\documentstyle[12pt]{l-aa} 
\input epsf 

\def\gsim{\lower.4ex\hbox{$\;\buildrel >\over{\scriptstyle\sim}\;$}}
\def\lsim{\lower.4ex\hbox{$\;\buildrel <\over{\scriptstyle\sim}\;$}}

  \def\bib{\bibitem{}}
\newcommand{\xia}{\overline{\xi}}
\newcommand{\rhoa}{\overline{\rho}}
\newcommand{\rhob}{\overline{\rho}}
\newcommand{\nbar}{\overline{n}}
\newcommand{\gam}{\gamma}
\newcommand{\pl}{\partial}

\newcommand{\beq}{\begin{equation}}
\newcommand{\eeq}{\end{equation}}
\begin{document}
%
% US additions
% 
\topmargin=2.5 cm

\thesaurus{Sect.02 (12.12.1; 11.05.2; 11.17.3; 11.09.3)} 
\title{The reheating and reionization history of the universe.}
\author{Patrick Valageas\inst{1,}\inst{2} \and Joseph Silk\inst{2,}\inst{3}}
\institute{Service de Physique Th\'eorique, CEA Saclay, 91191 
Gif-sur-Yvette, France 
\and
Center for Particle Astrophysics, Department of Astronomy and Physics,
University of California, Berkeley, CA 94720-7304, USA
\and
Institut d'Astrophysique de Paris, CNRS, 98bis Boulevard Arago,
F-75014 Paris, France}
\date{Received ;  }
\maketitle 
\markboth{P. Valageas \& J. Silk: Reheating and reionization}{P. Valageas 
\& J. Silk: Reheating and reionization}

\begin{abstract}

We incorporate quasars into an analytic model to describe the
reheating and reionization of the universe.  In combination with a
previous study of galaxies and Lyman-$\alpha$ clouds, we are able to
provide a unified description of structure formation, verified against
a large variety of observations. We also take into account the
clumping of the baryonic gas in addition to the presence of collapsed
objects. 

We consider two cosmologies: a critical universe with a CDM
power-spectrum and an open universe with $\Omega_0=0.3$, $\Lambda=0$. 
The derived quasar luminosity function agrees reasonably well with
observations at $z<4.5$ and with constraints over larger redshifts
from the HDF. The radiation produced by these objects at $z \sim 16$
slowly reheats the universe which  gets suddenly reionized
at $z_{ri}=6.8$ for the open universe ($z_{ri}=5.6$ for the critical
density universe). The UV background radiation simultaneously increases
sharply to reach a maximum of $J_{21} = 0.18$ at $z=2.6$, but
shows strong ionization edges until $z \leq 1$. The metallicity of the
gas increases quickly at high $z$ and is already larger than $0.01
Z_{\odot}$ at $z=10$. The QSO number counts and the helium opacity
constrain the reionization redshift to be  $z_{ri} \sim
6$. We confirm that a population of faint quasars is needed in order to
satisfy the observations. Due to the low reionization redshift, the damping
of CMB fluctuations is quite small, but future observations (e.g. with
the NGST) of the multiplicity functions of radiation sources and of
the HI and HeII opacities will strongly constrain  scenarios in which
reionization is due to QSOs. The reasonable agreement of our
results with observations (for galaxies, quasars, Lyman-$\alpha$
clouds and reionization constraints) suggests that such a model should be fairly realistic.

\end{abstract}


\section{Introduction}

An important goal of cosmology is to describe the structure formation
processes which led to the wide variety of astrophysical objects we
observe in the present universe, from Lyman-$\alpha$ clouds to
galaxies and clusters. Several studies have shown that the usual
hierarchical scenarios (like the standard CDM model) can provide
predictions which agree reasonably well with observations for galaxies
as well as for Lyman-$\alpha$ clouds. This corresponds to objects at
$z \leq 5$. However, it is possible to constrain the earlier evolution
of the universe by studying the reheating and reionization history
implied by such models. Indeed, observations show that the universe is
highly photo-ionized by $z=5$ and a large reionization redshift could
imprint a  signature on the CMB radiation. Moreover, future
missions such as  NGST could for instance detect quasars at high
redshifts $z>5$.

In this article, we present an analytic model for the
reheating and reionization history of the universe, adopting a CDM
power spectrum in a critical density and in an open universe. Similar studies
have been performed previously via numerical simulations
(e.g. Gnedin \& Ostriker 1997) and analytic approaches (e.g. Haiman
\& Loeb 1997; Haiman \& Loeb 1998) based on the Press-Schechter
prescription (Press \& Schechter 1974). However, previous analytic
models were often developed for this specific purpose (i.e. they were not
derived from a model already checked in detail against observations
of galaxies or Lyman-$\alpha$ clouds) and neglected the clumping of
the gas (except for the presence of virialized objects used to count
galaxies). Thus, the main motivations of our present study are to:

- describe these early stages of structure formation through a 
self-consistent model
which has already been applied to galaxies (Valageas \& Schaeffer
1998) and to Lyman-$\alpha$ clouds (Valageas et al.1999a).

- take into account  the broad range of
density fluctuations within the IGM
through our description of Lyman-$\alpha$ clouds. 

- use this feature to constrain our model against several
observations: notably the QSO number counts and the Gunn-Peterson test
(for HI and HeII).

- develop a simple analytic model which can predict many properties
of the universe (galaxy and quasar luminosity functions, temperature
and ionization state of the IGM, intensity and spectrum of the UV
background radiation and  fraction of matter within stars) and provide
a complementary tool  to numerical simulations.

Consideration of the various objects involved in our work (beyond
the just-virialized halos which are usually studied) is made possible
because of a specific description of the density field based on the
assumption that the many-body correlation functions obey the scaling
model detailed in Balian \& Schaeffer (1989) and checked numerically
in Colombi et al.(1997). This allows one to define the
various mass functions of interest, as described in Valageas \&
Schaeffer (1997; also in Valageas et al.1999b), and to go beyond the
scope of the usual Press-Schechter approximation (Press \& Schechter
1974). The main advantage of our approach is thus to provide a globally
consistent picture of structure formation in the universe, within the
framework of a hierarchical scenario.

This article is organized as follows. In Sect.\ref{Multiplicity
functions} we describe our prescription for mass functions. Next, in
Sect.\ref{Galaxy formation} we review our model for galaxy formation,
described in more detail in Valageas \& Schaeffer (1998) while in
Sect.\ref{Quasar radiative output} we deal with our prescription for
quasars. In Sect.\ref{Lyman-alpha clouds} we summarize 
the relevant aspects of our model
for Lyman-$\alpha$ clouds (Valageas
et al. 1999a).  We describe the calculation of the evolution of the
IGM properties (temperature, UV background radiation, ionization
state) in Sect.\ref{Evolution of the IGM} and in 
Sect.\ref{Opacity}. Finally, in Sect.\ref{Open universe} and Sect.\ref{Critical universe} we present our results for the case of an open universe and then for a critical universe.

\section{Multiplicity functions}
\label{Multiplicity functions}

We first review  the method we use to obtain the mass
functions of various astrophysical objects, specifically galaxies and
Lyman-$\alpha$ clouds.  We consider objects of dark matter mass $M$ to
be defined by a density threshold $\Delta(M,z)$. This constraint
depends on the class of astrophysical objects one considers and
it allows us for instance to distinguish clusters from galaxies which
correspond to higher density contrasts (see VS II). Lyman-$\alpha$
clouds are also formed by several populations of different objects
which are not always defined by a constant density threshold (see
Sect.\ref{Lyman-alpha clouds}). Note that
such a goal is beyond the reach of the usual Press-Schechter
prescription (Press \& Schechter 1974) which only deals with
``just-collapsed halos'' while we wish to describe simultaneously a
wide variety of objects. In any case, we attach to each halo a
parameter $x$ defined by:
\beq
x(M,z) =  \frac{1+\Delta(M,z)}{\; \xia[R(M,z),z] \;}
\label{xnl}
\eeq
where 
\[
\xia(R) =   \int_V \frac{d^3r_1 \; d^3r_2}{V^2} \; \xi_2 ({\bf r}_1,{\bf r}_2) 
 \;\;\;\;\; \mbox{with} \;\;\;\;\; V= \frac{4}{3} \pi R^3
\]
is the average of the two-body correlation function $\xi_2 ({\bf
r}_1,{\bf r}_2)$ over a spherical cell of radius $R$ and provides the
measure of the density fluctuations in such a cell. Then, we write the
multiplicity function of these objects (defined by the constraint
$\Delta(M,z)$) as (see VS I):
\beq
\eta(M,z) \frac{dM}{M}  = \frac{\rhob}{M} \; x^2 H(x) \; \frac{dx}{x}    
 \label{etah}
\eeq
where $\rhob$ is the mean density of the universe at redshift $z$,
while the mass fraction in halos of mass between $M$ and $M+dM$ is:
\beq
\mu(M,z) \frac{dM}{M} = x^2 H(x) \; \frac{dx}{x}     \label{muh}
\eeq
The scaling function $H(x)$ depends only on the initial spectrum of
the density fluctuations and must be obtained from numerical
simulations. However, from theoretical arguments (see VS I and Balian
\& Schaeffer 1989) it is expected to follow the asymptotic behaviour:
\[
x \ll 1 \; : \; H(x) \propto x^{\omega-2} \hspace{0.3cm} , \hspace{0.3cm}
 x \gg 1 \; : \; H(x) \propto x^{\omega_s-1} \; e^{-x/x_*}
\]
with $\omega \simeq 0.5$, $\omega_s \sim -3/2$, $x_* \sim 10$ to 20
and by definition it must satisfy
\beq
\int_0^{\infty}  x \; H(x) \; dx   =  1 
\eeq
The correlation function $\xia$, that measures the non-linear
fluctuations at scale $R$ can be modelled in a way that accurately
follows the numerical simulation. The
mass functions obtained from (\ref{etah}) for various constraints
$\Delta(M)$ were checked against the results of numerical simulations
in Valageas et al.(1999b) in the case of a critical universe with an
initial power-spectrum which is a power-law: $P(k) \propto k^n$ with
$n=0, -1$ and $-2$. This study showed that this model provides a
reasonable approximation to the mass functions obtained in the
simulations and that it works quite well for the two cases we shall
need in the present article: i) a constant density threshold $\Delta
\sim 178$ and ii) a constant radius constraint (or
$(1+\Delta) \propto M$). Moreover, the results of Valageas et
al.(1999b) showed that $H(x)$ is close to a similar scaling function
$h(x)$ obtained from the counts-in-cells statistics, as expected from
theoretical considerations (see for instance VS I).

It is clear that the model outlined above provides a unified
description of various astrophysical objects which are obtained from
the same non-linear density field. This is a great advantage of this
approach since it ensures that we can model a wide variety of objects,
from low density Lyman-$\alpha$ clouds to high density bright
galaxies, in a fully consistent way. Then, we can study the interplay
between these various structures as they develop progressively.

\section{Galaxy formation}
\label{Galaxy formation}

In this paper we wish to study the reionization history of the
universe. Since a large part of the ionizing radiation will be emitted
by stars, we first need to devise a model for galaxy and star
formation. We shall use a simplified version of the model described in
detail in VS II and which was there compared with many observations.
 One can define galaxies by the requirement
that two constraints be satisfied by the underlying dark matter halo:
1) {\it a virialization condition} $\Delta > \Delta_c$ (where
$\Delta_c(z) \sim 178$ is given by the spherical model and is constant
for a critical universe) and 2) {\it a cooling constraint} $t_{cool} <
t_{H}$ which states that the gas must have been able to cool within a
few Hubble times at formation. However, at high redshifts $z>1$ the
cooling constraint becomes irrelevant since any object which satisfies
1) also satisfies 2). Hence since we are mainly interested in large
redshifts $z>1$ we shall simply define galaxies by the virialization
condition $\Delta = \Delta_c$. We also require that the virial
temperature $T$ of the halo be larger than the ``cooling temperature''
$T_{cool}(z)$ at redshift $z$. The latter corresponds to the smallest
virialized objects which can cool efficiently at redshift $z$, defined
by the constraint:
\beq
t_{cool} = s \; t_H  \label{tcool}
\eeq
where $s=6$ is a proportionality factor (one must have $s>1$ since
cooling is more efficient within the halo where the density is larger
than on its boundary and cooling accelerates as baryons collapse). Here
$t_{H}(z)$ is the age of the universe at redshift $z$ while $t_{cool}$
is the cooling time of a halo with density contrast $\Delta_c(z)$,
mass $M$, taking into account both cooling (recombination, molecular
cooling) and heating (by the background UV flux) processes. Since
the physical properties of virialized halos with temperature $T$ and
density contrast $\Delta_c(z)$ are different from the IGM, we let the
chemical reactions (involving HI, HII, H$^-$, H$_2$, H$_2^+$, HeI,
HeII, HeIII and e$^-$) evolve for a Hubble time $t_H$ within this
environment (defined by $T$ and $\Delta_c$) before we evaluate the
cooling time $t_{cool}$. The main effect is that at large redshifts
such clouds may produce enough molecular hydrogen to make molecular
cooling efficient while with the use of the IGM abundances one would
underestimate this contribution; see for instance Tegmark et al.(1997)
for a detailed discussion. This will also appear clearly below in
Fig.\ref{figtcoolO03} where we compare the main contributions to
cooling for both the IGM and these cooling halos. The virial
temperature $T_{cool}$ also defines the mass $M_{cool}(z)$ and the
radius $R_{cool}(z)$ of the smallest objects which can cool and
eventually form stars at redshift $z$. From the lower-bound
$T_{cool}(z)$ and the virialization constraint $\Delta=\Delta_c(z)$, we
obtain the mass function of galaxies at redshift $z$ using
(\ref{etah}).

Next, we must attach a specific stellar content to these galactic
halos. We shall again use the star formation model described in VS
II. This involves 4 components: (1) short-lived stars which are
recycled, (2) long-lived stars which are not recycled, (3) a central
gaseous component which is deplenished by star formation and ejection
by supernovae winds, and replenished by infall from (4) a diffuse gaseous
component. The star-formation rate $dM_s/dt$ is proportional to the
mass of central gas with a time-scale set by the dynamical time. The
mass of gas ejected by supernovae is proportional to the
star-formation rate and decreases for deep potential wells as $1/T$,
in a fashion similar to that adopted by Kauffmann et al.(1993). It was seen in
VS II that for such a model a good approximation for the
star-formation rate is:
\beq
\frac{dM_s}{dt} = \frac{M_g}{\tau_0} \hspace{0.5cm} \mbox{with} \hspace{0.5cm}
\tau_0 \simeq \left( 1+\frac{T_{SN}}{T} \right) \; \tau_d   \label{SFR}
\eeq
where $M_g$ is the total mass of gas, $\tau_d$ is the dynamical
time and $T_{SN}=10^6$ K describes the ejection of gas by supernovae and stellar winds: 
\beq
T_{SN} =  \frac{2 \; \epsilon \; E_{SN} \; \mu m_p \; \eta_{SN}}{3 \; k \; m_{SN}} \sim 10^{6} \; \mbox{K}  
\label{T0SN}
\eeq
Here $\epsilon \sim 0.1$ is the fraction of the energy $E_{SN}$ delivered by supernovae transmitted to the gas ($E_{SN} = 10^{51}$ erg) while $\eta_{SN}/m_{SN} \simeq 0.005 \; M_{\odot}^{-1}$ is the number of supernovae per solar mass of stars formed. Note that for halos defined by a constant density threshold $\Delta_c \sim 178$ we have $\tau_d \sim t_H(z)$. Although (\ref{SFR}) was obtained for small galaxies with $T \ll T_{SN}$ (which is the
range we are mainly interested in) it also provides a reasonable approximation for large galaxies $T > T_{SN}$. In the case $\Omega=1$ we obtain in our model for a galaxy similar to the Milky Way (i.e. with a circular velocity $V_c=220$ km/s): $\tau_0 \simeq 7 \; 10^9$ years and $dM_s/dt \simeq 5 M_{\odot}/$year (see VS II). This star formation rate is consistent with observations (McKee 1989). Then the mass of gas at time $t$ within the galaxy is given by:
\beq
M_g = M_{g0} \; e^{-t/\tau_0}  \label{Mg}
\eeq
where $M_{g0}=M_b$ is the initial mass of baryons which we take to be
proportional to the dark matter mass $M$:
\beq
M_b = \frac{\Omega_b}{\Omega_0} \; M
\eeq
From this model, the star-formation rate per Mpc$^3$ is:
\beq
\begin{array}{ll}
{\displaystyle \left( \frac{d\rho_s}{dt} \right) = } & {\displaystyle
\frac{\Omega_b}{\Omega_0} \; \frac{\rhob(z)}{t_H} \; \int_{x_{cool}}^{\infty} 
\frac{p}{\beta_d} \left(1+\frac{T_{SN}}{T} \right)^{-1}   }   
\\ & \\
& {\displaystyle  \hspace{1cm} \times \; e^{ - \frac{p}{\beta_d} 
(1+\frac{T_{SN}}{T})^{-1} } \; x^2 H(x) \; \frac{dx}{x}  }
\end{array}
\label{SFRav}
\eeq
where $p/\beta_d = 0.7$ is a parameter of order unity which enters the
definition of the dynamical time $\tau_d$. The significance of each
term in this expression is clear and the temperature dependence simply
states that the average star formation efficiency of small galaxies is
small as the gas is easily expelled by supernovae. Note that in the
original model described in VS II for bright galaxies at low redshifts,
the star formation rate declines since most of the gas has already
been consumed. This does not appear in (\ref{SFRav}) because we defined
all galaxies by $\Delta=\Delta_c$ while at low $z$ for large $T$ the
cooling constraint implies that $(1+\Delta) \propto M$ (i.e. $R$ is
constant) which decreases the galactic dynamical time and increases
the ratio $t/\tau_0$ which enters (\ref{Mg}).

To derive the radiation emitted by galaxies, we do not need their
global star formation rate but their stellar content. However, as
shown in VS II, the mass in the form of short-lived stars (i.e. with a
life-time $\tau_{sh}$ small  compared to $t_H$) of mass $m$ to
$m+dm$ is given by:
\beq
dM_{sh} = d\eta \; \frac{\tau_{sh}}{\tau_0} \; M_g = d\eta \; \tau_{sh} \; 
\left( \frac{dM_s}{dt} \right)  \label{Msh}
\eeq
where $d\eta=m \phi(m) dm$ is the fraction of mass which goes into
such stars for each unit mass of stars which are formed. This depends on
the initial stellar mass function (IMF) $\phi(m)$. Since the stellar
radiation output at high energy ($\nu > 13.6$ eV) is dominated by the
most massive stars, the relation (\ref{Msh}) will be sufficient for our
purposes. Next, if we assume that stars radiate as blackbodies with
an effective temperature $T_{eff} \propto L^{0.13}$ and we use the
mean scalings $\tau_{sh} \propto m/L$ and $L \propto m^{3.3}$ we
obtain the energy output of such galaxies:
\beq
\left( \frac{\pl^2 E}{\pl \nu \pl t} \right)_s = \left( \frac{dM_s}{dt} 
\right) \; \frac{1 \mbox{yr}}{1 M_{\odot}} \; L_{\nu s}(\nu)
\label{dEdnudts}
\eeq
with
\[
L_{\nu s} (\nu) = \frac{10^{10} L_{\odot}}{\nu} \; \int m \phi(m) dm 
\frac{2\pi h \nu^4}{\sigma T^4 c^2 ( e^{h\nu/kT}-1)}
\]
From the radiation emitted by individual galaxies we now wish to
estimate the energy received by a random point in the IGM. We shall
write the source term $S_{\nu s}$ due to stellar radiation for the
background UV flux $J_{\nu}$, see (\ref{Jnu}), as the following
average:
\beq
S_{\nu s} = \frac{c}{4 \pi} \; \int \eta_g(x) \frac{dx}{x} \; \left( 
\frac{\pl^2 E}{\pl \nu \pl t} \right)_s (x) \; e^{-\tau_s(x)}  \label{Snus}
\eeq
where $\eta_g(x) dx/x$ is the mass function of galaxies, obtained from
(\ref{etah}) as described previously, while $\tau_s$ is a mean opacity
which takes into account the fact that the radiation emitted by
galaxies can be absorbed by the IGM {\it and} Lyman-$\alpha$
clouds. We shall come back to this term later. Thus, we get in this
way a simple model for the stellar radiative output from our more
detailed description of galaxy formation. The reader is referred to VS II
for a more precise account
of the details and predictions of our galaxy formation  model. Note
that
our
prescription is consistent with such  observations as the
Tully-Fisher relation and  the B-band luminosity function.

\section{Quasar radiative output}
\label{Quasar radiative output}

In addition to galaxies we also need to describe the radiation emitted
by quasars which provide a non-negligible contribution to the
background radiation field, especially at the high frequencies $\nu >
24.6$ eV which are relevant for helium ionization. We shall again
follow the formalism of
VS II to obtain the quasar luminosity function, in a fashion
similar to Efstathiou \& Rees (1988) and Nusser \& Silk (1993). We
assume that the quasar mass $M_Q$ is proportional to the mass of gas
$M_{gc}$ available in the inner parts of the galaxy: $M_Q = F \;
M_{gc}$. Note that for galaxies which have not yet converted most of their gas
into stars (i.e. all galaxies except those with $T > T_{SN}$ at $z <
1$) this also implies $M_Q \sim F \; M_s$ where $M_s$ is the stellar
mass. Indeed, for $t_H < \tau_0$ (where $t_H$ is the age of the universe) we have:
\beq
M_s \sim t_H/\tau_0 \; M_g
\eeq by definition of $\tau_0$, see (\ref{SFR}), while the mass
$M_{gc}$ of cold central gas $M_{gc}$ satisfies: \beq M_{gc} \sim
\left( 1+\frac{T_{SN}}{T} \right)^{-1} \; M_g
\label{Mgc}
\eeq The factor $(1+T_{SN}/T)$ translates the fact that in our model
supernovae eject part of the star-forming gas out of the galactic
center into the larger dark matter halo (VS II). Hence
we get $M_{gc} \sim M_s$. Of course, at late times for bright galaxies
when most of the gas has been consumed we have $M_{gc} \ll M_s$. Then
the mass of gas available to feed the quasar declines with time. This
leads to a high luminosity cut-off at low $z$ for the quasar
luminosity function since in this regime very massive galaxies have
less gas than smaller ones which underwent less efficient star
formation (see VS II and Sect.\ref{Quasar luminosity function}). We
shall use $F=0.01$ for $\Omega=1$ and $F=0.006$ for
$\Omega_0=0.3$. Note that observations (Magorrian et al.1998) find
$M_Q \simeq 0.006 \; M_s$ in large galaxies. Next we write the
bolometric luminosity $L_Q$ of the quasar as: \beq L_Q =
\frac{\epsilon \; M_Q \; c^2}{t_Q} \eeq where $\epsilon = 0.1$ is the
quasar radiative efficiency (fraction of central rest mass energy
converted into radiation) and $t_Q$ is the quasar life-time. Since we
shall assume that quasars radiate at the Eddington limit we have: $t_Q
= 4.4 \; \epsilon \; 10^8$ yr. Thus, the quasar luminosity attached to
a galaxy of dark matter mass $M$, virial temperature $T$, is: \beq L_Q
= \frac{\epsilon \; F}{t_Q} \; \frac{\Omega_b}{\Omega_0} \; \left( 1 +
\frac{T_{SN}}{T} \right)^{-1} \; M c^2 \label{LQ} \eeq
As seen above in (\ref{Mgc}), the temperature term comes from the fact that in our galactic model,
small galaxies ($T \ll T_{SN}$) are strongly influenced by supernova
feedback which expells part of their baryonic content from  the
inner regions. Note however that this term does not enter the relation
(quasar mass) - (stellar mass) as it cancels out on both sides. Next
we obtain the quasar multiplicity function from the galaxy mass
function as:
\beq
\eta_Q(M_Q) \frac{dM_Q}{M_Q} = \lambda_Q \; \eta_g(M) \frac{dM}{M} \; 
\mbox{Min} \left[ 1 , \frac{t_Q}{t_M} \right]   \label{etaQ}
\eeq
The factor $\lambda_Q < 1$ (we use $\lambda_Q=0.06$) is the fraction of galactic halos
which actually harbour a quasar while $t_M$ is the evolution
time-scale of galactic halos of mass $M$ defined by:
\beq
t_M^{-1} = \frac{1}{\eta_g(M)} \; \frac{\pl}{\pl t} \eta_g(M)
\eeq
Since the quasar life-time $t_Q \sim 10^8$ yr is quite short, this
reduces to $\eta_Q(M_Q) dM_Q/M_Q = \lambda_Q \; t_Q \; \pl
\eta_g / \pl t \; dM/M$. Together with (\ref{LQ}) the relation
(\ref{etaQ}) provides the quasar luminosity function. Thus, we only
have two parameters: $(\epsilon \; F/t_Q)$ (which only depends on $F$,
constrained by the observed (quasar mass)/(stellar mass) ratio, for
quasars shining at the Eddington luminosity) which enters the
mass-luminosity relation, and $(\lambda_Q \; t_Q)$ which appears as a
simple normalization factor in the luminosity function. Hence a larger
fraction of quasars $\lambda_Q$  together
with a smaller life-time $t_Q$ would give the same results, so that we could also choose $\lambda_Q=1$. In a fashion similar to what we did
for galaxies we can now derive the quasar radiative output. We first
write the radiation emitted by an individual quasar as:
\beq
\left( \frac{\pl^2 E}{\pl \nu \pl t} \right)_Q = \frac{L_Q}{\nu_B} \; \left( 
\frac{L_B}{L_{bol}} \right) \; \left( \frac{\nu_B}{\nu} 
\right)^{\alpha}
\eeq
where $(L_B/L_{bol}) = 0.094$ is the conversion factor from bolometric
luminosity to B-band luminosity ($L_B = \nu_B L_{\nu}(\nu_B)$ at
$\nu_B=2.8$ eV), taken from Elvis et al.(1994), while $\alpha=1.5$ is
the local slope of the quasar spectrum. Then, the source term $S_{\nu
Q}$ for the background radiation due to quasars is:
\beq
S_{\nu Q} = \frac{c}{4 \pi} \; \int \eta_Q(x) \frac{dx}{x} \; \left( 
\frac{\pl^2 E}{\pl \nu \pl t} \right)_Q (x) \; e^{-\tau_Q(x)}  \label{SnuQ}
\eeq
where again $\tau_Q$ is a mean opacity which we shall describe later.

\section{Lyman-$\alpha$ clouds}
\label{Lyman-alpha clouds}

The description of gravitational clustering used in this article
allows one to build a model for Lyman-$\alpha$ clouds 
(Valageas et al.1999a). We shall take advantage of this
possibility to include these objects in the present study. Indeed,
although at high redshifts they do not contribute significantly to the
total opacity (which comes mainly from the uniform component of the
IGM) since only a small fraction of baryonic matter has been allowed to
form bound objects, at redshifts close to the reionization epoch they
already provide a non-negligible opacity.
We identify Lyman-$\alpha$ absorbers as three different classes of
objects, which we shall briefly describe below.

\subsection{Lyman-$\alpha$ forest}

We assume that after reionization the gas within low-density halos is
reheated by the UV flux to a temperature $T_{Ly} = 3 \; 10^4$ K. Hence
in such shallow potential wells, baryonic density fluctuations are
erased over scales $R_{dLy}$ defined as in (\ref{Rd}) but with the
temperature $T_{Ly}$. This builds our first class of objects defined
by their radius $R_{dLy}(z)$ and virial temperatures $T < T_{Ly}$. The
multiplicity function of these mass condensations is again obtained
from (\ref{etah}). The fraction of neutral hydrogen at low $z$ is
evaluated by assuming photo-ionization equilibrium. At high $z$ prior
to reionization, when the UV flux is very small and cannot heat the
gas, we simply take $T_{Ly}=T_{IGM}$ while the fraction of neutral
hydrogen is unity. Since the baryonic density is roughly uniform
within these objects (by definition) we consider that each halo
produces one specific mean column density on any intersecting
line-of-sight (we neglect the small dependence on the impact parameter
due to geometry). At low $z$ this population can be identified with
the Lyman-$\alpha$ forest. Note that, as explained in details in Valageas et al.(1999a), our approach is also valid for clouds which are not spherical objects of radius $R_{dLy}$ but filaments of thickness $R_{dLy}$ and length $L \gg R_{dLy}$. This is due to the growth of the density fluctuations on smaller scales (along with $\xia$) and to the direction jumps of filamentary structures.

Here we note that models for the Lyman-$\alpha$ forest are often
classified in two categories: 1) mini-halo models and 2) IGM density
fluctuations. In case 1), one considers that Lyman-$\alpha$ absorbers
are discrete clouds formed by bound collapsed objects (or halos
confined by the IGM pressure) which occupy a small fraction of the
volume. On the other hand, in case 2) (which is currently favored) one
assumes that absorption comes from a continuous medium (the IGM) with
relatively small density fluctuations. Although in our model we
identify distinct patches of matter (of size $R_{dLy}$) as in 1), the
underlying picture corresponds to the case 2). Indeed, as we consider
regions with an ``overdensity'' $(1+\Delta)$ from $\sim 20$ down to
$(1+\Delta)_{IGM}$, defined below in (\ref{DeltaIGM}), which can be as
low as $10^{-3}$, see Fig.\ref{figclumpO03}, we take into account {\it
all the volume} of the universe. Hence our Lyman-$\alpha$ forest
absorbers are made of a broad range of density fluctuations within the IGM which fill all the space between galactic halos (which we describe below as they
form Lyman-limit and damped systems and only occupy a negligible
fraction of the volume, as seen in Fig.\ref{figFracO03}). Note that
this would not be possible if we were to consider density fluctuations
defined by a constant density threshold $(1+\Delta)_{th} > 1$ since
this would imply that we probe at most a fraction $1/(1+\Delta)_{th}$
of the volume of the universe. We identify the lowest density regions (i.e. with a density contrast $\Delta_{IGM}$), which are also the most numerous and fill most of the volume, with the IGM. A patch of matter with this density would only make up a column density $N_{HI} \sim 10^6$ cm$^{-2}$ on a scale $R_{dLy}$ at $z=0$.

\subsection{Lyman-limit systems}

Potential wells with a large virial temperature $T > T_{Ly}$ do not
see their baryonic density profile smoothed out and they also retain
their individuality. Thus, we define a second class of objects
identified to the ionized outer shells of virialized halos,
characterized by their density contrast $\Delta_c$ and satisfying $T >
T_{Ly}$. The deepest of these potential wells (such that $T >
T_{cool}$) corresponds to the galactic halo described in
Sect.\ref{Galaxy formation}. We assume that the mean density profile is a
power-law $\rho \propto r^{-\gam}$ (with $\gam=1.8$) so that each
object can now produce a broad range of absorption lines, as a function of
the impact parameter of the line-of-sight. This population can be
identified with the Lyman-limit systems.

\subsection{Damped systems}

The deep cores of the virialized halos described above which are not
ionized because of self-shielding (at low $z$) form our third
population of objects. One halo can again produce various absorption
lines for different impact parameters. At high $z$, prior to
reionization, halos are entirely neutral so that the previous
contribution of ionized shells disappears and we only have two classes
of objects: these neutral virialized halos and the ``forest'' objects.

\section{Evolution of the IGM}
\label{Evolution of the IGM}

 We now turn to
the IGM itself. We model the universe at a given redshift $z$ as
a uniform medium (the IGM), characterized by a density contrast
$\Delta_{IGM}$, a gas temperature $T_{IGM}$ and a background radiation
field $J_{\nu}$, which contains some mass condensations recognized as
individual objects identified with galaxies or Lyman-$\alpha$ clouds
as  described above.

Since the gas in the IGM has non-zero temperature $T_{IGM}$, baryonic
density fluctuations are erased over scales of order $R_{d}(z)$ within
shallow potential wells with a virial temperature $T_{vir}<T_{IGM}$ or
within ``voids'', with: 
\beq 
R_d(z) = \frac{1}{2} \; t_H \; C_s =
\frac{1}{2} \; t_H \; \sqrt{ \frac{\gam k T_{IGM}}{\mu m_p} }
\label{Rd}
\eeq 
where $C_s$ is the sound speed, $t_H$ the age of the universe,
$m_p$ the proton mass and $\gam \sim 5/3$. Indeed, the pressure
dominates over gravitation for objects such that $T_{vir}<T$. Note
that the damping scale $R_d$ is different from the Jeans scale: 
\beq
R_{J}(z) = \sqrt{ \frac{\gam k T_{IGM}}{4 \pi {\cal G} \mu m_p
\rho_{DM}} }
\label{RJ}
\eeq 
Both scales are equal (up to a normalization factor of order
unity) if the dark matter density is equal to the mean universe
density: $\rho_{DM}=\rhoa$. However, we shall consider underdense
regions where $(1+\Delta)$ can be as low as $10^{-3}$, see
Fig.\ref{figclumpO03} below. Indeed, as an increasingly large
proportion of the matter content of the universe gets embedded within
collapsed objects as time goes on the density of the IGM (the volume
between these mass condensations) becomes much smaller than the mean
universe density. In this case where $\rho_{DM} < \rhoa$ we have $R_d
< R_J$. We use $R_d$ because of the finite age of the universe: the
medium cannot be homogenized over scales larger than those reached by
acoustic waves over the time $t_H$ (the scale $R_J$ corresponds to the
limit of large times). Note that for Lyman-$\alpha$ clouds we also use
$R_d$ as the characteristic scale, with $T_{Ly} = 3 \; 10^4$ K, since
we consider regions with very low or moderate densities $(1+\Delta) <
45$, see Sect.\ref{Lyman-alpha clouds} and Valageas et
al.(1999a). Then, the density contrast of the IGM is given by: 
\beq
(1+\Delta)_{IGM} = \mbox{Min} \left[ \;1 \; , \;
\xia(R_d)^{-\omega/(1-\omega)} \; \right]
\label{DeltaIGM}
\eeq 
This simply states that at high $z$ (when $\xia(R_d) \ll 1$) we
have $\rho_{IGM} = \rhoa$ (i.e. the universe is almost exactly a
uniform medium on scale $R_d$) while at low $z$ we have $\rho_{IGM} <
\rhoa$ since most of the matter is now within overdense bound
collapsed objects (clusters, filaments etc.) while most of the volume
(which we call the IGM) is formed by underdense regions.

Since the mean density of the universe is $<\rho>=\rhoa$ we define a
baryonic clumping factor $C_b = <\rho_b^2>/<\rho_b>^2$ by: 
\beq
\begin{array}{ll} 
C_b & = {\displaystyle F_{IGM,vol} \; (1+\Delta)_{IGM}^2 \; + \int
(1+\Delta) \; x^2 H(x) \frac{dx}{x} } \\ & \\ & \simeq
(1+\Delta)_{IGM}^2 + F_{Ly} <1+\Delta>_{Ly} + F_{vir} (1+\Delta_c)
\end{array}  
\label{Cb}
\eeq 
where we used the fact that the volume fraction $F_{IGM,vol}$
occupied by the IGM is very close to unity. Here $F_{Ly}$ and
$F_{vir}$ are the fractions of mass formed by Lyman-$\alpha$ forest
clouds (with a density contrast lower than $\Delta_c$) and by
virialized objects. Note that $C_b$ somewhat underestimates the actual
clumping of the gas since we did not take into account the collapse of
baryons due to cooling nor the slope of the density profile within
virialized halos. However, these latter characteristics are included
in our model for Lyman-$\alpha$ clouds. We also define the mean
density due to objects which do not cool as: 
\beq 
<1+\Delta>_n =
(1+\Delta)_{IGM} + \int_0^{x_{cool}} x^2 H(x) \frac{dx}{x}
\label{rhon}
\eeq 
Before reionization this corresponds to the density field of
neutral hydrogen since galactic halos (i.e. massive potential wells
with $x>x_{cool}$ which can cool) ionize most of their gas because of
the radiation emitted by their stars or their central quasar. We
obtain the mean square density in a similar fashion:
\[
<(1+\Delta)^2>_n = (1+\Delta)_{IGM}^2 + \int_0^{x_{cool}} (1+\Delta)
\; x^2 H(x) \frac{dx}{x}
\]
and the corresponding clumping factor is simply: 
\beq 
C_n =
\frac{<(1+\Delta)^2>_n}{<1+\Delta>_n^2}
\label{Cn}
\eeq 
The quantities $<1+\Delta>_n$ and $C_n$ characterize the density
fluctuations of neutral hydrogen within the IGM. Note that most of the
volume is occupied by regions which satisfy $(1+\Delta) \sim
(1+\Delta)_{IGM}$.

The gas which is within the IGM is heated by the UV background
radiation while it cools due to the expansion of the universe and to
various radiative cooling processes. Note that we neglect here the
possible heating of the IGM by supernovae. However supernova feedback
is included in our model for galaxy formation: we simply assume it
only affects the immediate neighbourhood of these galaxies (see also
McLow \& Ferrara 1998). Thus, we write for the evolution of the
temperature of the IGM: 
\beq 
\frac{dT_{IGM}}{dt} = - 2 \;
\frac{\dot{a}}{a} \; T_{IGM} \; - \; \frac{T_{IGM}}{t_{cool}} \; + \;
\frac{T_{IGM}}{t_{heat}} 
\label{TIGM} 
\eeq 
where $a(t)$ is the scale
factor (which enters the term describing adiabatic cooling due to the
expansion). The heating time-scale $t_{heat}$ is given by: \beq
t_{heat}^{-1} = \frac{4 \pi}{3/2 n_b k T_{IGM}} \; \sum_j \; \int n_j
\sigma_j(\nu) (\nu - \nu_j) J_{\nu} \frac{d\nu}{\nu} \label{theat}
\eeq where $j=$ (HI,HeI,HeII), $\nu_j$ is the ionization threshold of
the corresponding species, $n_j$ its number density in the IGM and
$n_b$ the baryon number density. The cooling time-scale $t_{cool}$
describes collisional excitation, collisional ionization,
recombination, molecular hydrogen cooling, bremsstrahlung and Compton
cooling or heating (e.g. Anninos et al.1997). Next, we can write the
evolution equation for the background radiation field $J_{\nu}$: 
\beq
\frac{\pl J_{\nu}}{\pl t} = -3 \; \frac{\dot{a}}{a} \; J_{\nu} \; + \;
\frac{\dot{a}}{a} \; \nu \; \frac{\pl J_{\nu}}{\pl \nu} \; - \;
k_{\nu} J_{\nu} \; + S_{\nu s} + S_{\nu Q} \label{Jnu} 
\eeq 
The first two terms on the r.h.s. describe the effects of the expansion of the
universe, while the last two terms represent the radiation emitted by
stars and quasars which we obtained previously. The absorption
coefficient $k_{\nu}$ is written as: 
\beq 
k_{\nu} = \frac{c}{1
\mbox{Mpc}} \left( \tau_{\nu, IGM}^1 + \tau_{\nu, NHI}^1 + \tau_{\nu,
NHeI}^1 + \tau_{\nu, NHeII}^1 \right) 
\eeq 
where $\tau_{\nu, IGM}^1$
is the opacity at frequency $\nu$ of the IGM over a physical length of
1 Mpc, while $\tau_{\nu, N_j}$ corresponds to the contribution by
``Lyman-$\alpha$'' clouds (i.e. discrete mass condensations as opposed
to the uniform component which forms the IGM). Thus we have: 
\beq
\tau_{\nu, IGM}^1 = \left( \sum_j \sigma_j(\nu) n_j \right) 1
\mbox{Mpc} 
\eeq 
Note that in this study we consider the medium as
purely absorbing and we neglected the reprocessing of ionizing
photons. From the evolution of the IGM temperature and the background
radiation field we can also follow the chemistry of the gas within
this uniform component. More precisely we consider the following
species: HI, HII, H$^{-}$, H$_2$, H$_2^+$, HeI, HeII, HeIII and e$^-$
(see for instance Abel et al.1997 for rate coefficients). Thus we
obtain the reionization history of hydrogen and helium together with
the spectral shape of the background radiation $J_{\nu}$.

\section{Opacity}
\label{Opacity}

In the previous calculations, (\ref{Snus}) and (\ref{SnuQ}), where we
described the source terms for the radiation field within the IGM we
introduced opacity factors to model the absorption of the radiation
emitted by quasars and stars by the IGM and Lyman-$\alpha$ clouds. We
shall deal with these terms in this section.

We consider that each source (galaxy or quasar) active at a given
redshift $z$ ionizes its surroundings over a radius $R_i$
given by:
\beq
R_{i,HII} = \left[ \frac{3}{4 \pi \alpha C_n n_H^2} \; \left( 1 - 
e^{-\alpha C_n n_H t} \right) \; \frac{dN_{\gam}}{dt} \right]^{1/3}  
\label{Ri}
\eeq
where $\alpha(3 \; 10^4$ K$)$ is the recombination rate (within the ionized bubbles), $n_H$ the mean
number density of hydrogen obtained from (\ref{rhon}), $C_n$ the clumping factor from (\ref{Cn}), $(dN_{\gam}/dt)$
the emission rate of ionizing photons from the source and $t$ its
age.

We have $t \leq t_H$ so we neglected here the influence of the
expansion of the universe and the time-dependence of the source
luminosity over its age. We take $t = t_H$ for galaxies and $t = t_Q$
for quasars. Although this procedure is consistent with our
prescription for the quasar luminosity function (we assumed quasars to
shine at the Eddington luminosity on the time-scale $t_Q \ll t_H$ and
then to fade) we somewhat overestimate the radiative output of
galaxies since the galaxy luminosity function decreases with $z$ over
the time-scale $t_H$. However, the relation (\ref{Ri}) should still
provide a correct estimate of the magnitude of this effect. Note that $R_i$ is smaller than the usual ``Stromgren'' radius which corresponds to the limit $t \rightarrow \infty$ in (\ref{Ri}). Indeed, the exponential term in (\ref{Ri}) can also be written as $\exp(-t/t_{rec})$ which shows that at redshifts $z \ll 100$ where the recombination time is larger than the age of the source (which is smaller than $t_H$) the ionization front is smaller than the Stromgren radius. This effect was also described by Shapiro \& Giroux (1987). In addition, these authors took into account the expansion of the universe but assumed a fixed number of sources. Here since we consider sources with a life-time $t \leq t_H$ we neglect the influence of the expansion of the universe but we take into account the increasing number of galaxies and quasars. 

Next, we obtain the volume fraction $Q_{HII_{s,Q}}$ (i.e. the filling
factor) occupied by such ionized bubbles around galaxies or quasars
as: 
\beq 
Q_{HII_{s,Q}} = \int \eta_{bubble_{s,Q}} (x) \; \frac{dx}{x}
\; \frac{4\pi}{3} R_{i_{s,Q}}^3
\eeq
For bubbles ionized by stellar radiation we have $\eta_{bubble_s} (x) = \eta_g(x)$ where $\eta_g(x) dx/x$ is the mass function of galaxies while for quasars we write:
\beq
\eta_{bubble_Q} (x) = \lambda_Q \; \eta_g(x) \; \mbox{Min} \left[ 1 , \mbox{Max} \left( \frac{t_Q}{t_M} , \frac{t_{rec}}{t_M} \right) \right]
\label{etabubbleQ}
\eeq
where $t_{rec}$ is the recombination time within the ionized bubbles. This differs from the quasar mass function (\ref{etaQ}) through the term $t_{rec}/t_M$ because a region remains ionized over a time-scale $t_{rec}$ which may be longer than the quasar life-time $t_Q$. Note that our general procedure only provides an upper bound to the
actual efficiency of radiative processes since we did not include
absorption within the host galactic halo itself. We do not integrate
$Q_{HII_{s,Q}}$ over time since this is already done in (\ref{Ri}) and
the sharp rise with time of the luminosity functions (before
reionization) ensures that the radiative output is dominated by recent
epochs. Moreover, since we have $t_{rec} < t_H$, as shown below in Fig.\ref{figtionrecO03} (curve $t_{rec,bubble}$), ionized bubbles do not survive more than a Hubble time unless new sources (galaxies or quasars) appear. The filling factor within the IGM is written as the sum of the
contributions from galaxies and quasars:
\beq
Q_{HII,IGM} = Q_{HII_s} + Q_{HII_Q}
\eeq
Of course the previous considerations only apply to high redshifts
prior to reionization when the universe is almost completely
neutral. At reionization these bubbles overlap and the background UV
flux gets suddenly very large as absorption drops. At later times the
whole universe is ionized so there are no more discrete bubbles (and
formally $Q_{HII}=1$). We also define in a similar fashion the filling
factors $Q_{HeII}$ and $Q_{HeIII}$ which describe regions around
quasars where helium is singly or doubly ionized. Since the stellar
radiation shows an exponential decrease at high frequencies quasars
are the only relevant source for this process. The filling factor
$Q_{HII,IGM}$ obtained above will be used to obtain the IGM
opacity.

 However, as we explained previously we also consider the
universe to contain numerous clouds which contribute to the opacity
seen by the radiation field. We shall first consider that these clouds
are ionized (or more exactly that their number density drops
significantly) within the radius $R_{i,cl} = R_i$ defined in
(\ref{Ri}) from the quasar. In other words, most of the opacity comes
from clouds located deeply within the IGM where the background
radiation $J_{\nu}$ is very small (before reionization) since close to
quasars (within $R_{i,cl}$) the local radiation suddenly gets much
higher. Note that the distribution of Lyman-$\alpha$ clouds we
calculate is indeed obtained from the IGM background radiation, which
calls for the cutoff $R_{i,cl}$. However, at low $z$ after
reionization the ``sphere of influence'' of a quasar is no longer
given by $R_{i,cl}$ (since the whole medium is ionized). Instead, we
define the radius $R_{J,cl}$ by:
\beq
\frac{1}{4 \pi R_{J,cl}^2}  \; \left( \frac{dN_{\gam}}{dt} \right)_Q = 
10 \; \frac{J_{21}}{h} 
\eeq
where $h$ is Planck constant and $J_{21}$ is a measure of the background radiation within the IGM
in the ionizing part of the spectrum:
\beq
J_{21} = \frac{ \int J_{\nu} \; \sigma_{HI}(\nu) \; \frac{d\nu}{\nu} }
{ \int \sigma_{HI}(\nu) \; \frac{d\nu}{\nu} }  \label{J21}
\eeq
Thus, the ``sphere of influence'' of a quasar is defined by the region
of space around the source where the radiation emitted by this quasar
is significantly larger than the background radiation (at high $z$
when the medium is neutral this corresponds to $R_{i,cl}$, while at
low $z$ when the IGM is ionized this is given by $R_{J,cl}$). Thus, in
practice we shall simply use $R_{cl} =
\mbox{Min}[ R_{i,cl} , R_{J,cl} ]$ to obtain the volume fraction
$Q_{HII,cloud}$ where the number density of Lyman-$\alpha$ clouds is
significantly lower than within the IGM:
\beq
Q_{HII,cloud} = \int \eta_{bubble_Q} (x) \; \frac{dx}{x} \; \frac{4\pi}{3} 
R_{cl}^3 
\eeq

As we shall see from the numerical results, at low redshifts $z <
z_{ri}$ when the universe is reionized the opacity is very low (the UV
flux is large) so that absorption plays no role for the evolution of
$J_{\nu}$. Thus, in practice the radius $R_{J,cl}$ is irrelevant. It
only provides some information on the properties of the universe but
it does not influence the behaviour of the latter. Thus, while
$Q_{HII,IGM}$ increases with time until it reaches unity as the
universe gets reionized, $Q_{HII,cloud}$ will first grow before
reionization as the volume occupied by the ionized bubbles increases
and then decrease at $z \ll z_{ri}$ because the quasar luminosity
function drops at low $z$. Since a fraction of volume $Q$ translates
into the same fraction $Q$ along a random line of sight (neglecting
correlations in the distributions of sources) we write the opacity
$\tau_{\nu}(r)$ seen from a point in the IGM to a source located at
the distance $r$ as:
\beq
\tau_{\nu}(r) = \sum_j \left( \tau^1_{\nu,N_j} \; Q_{j,cloud} + 
\tau^1_{\nu,IGM_j} \; Q_{j,IGM} \right) \; \frac{r}{1 \mbox{Mpc}}  
\label{tauQHI}
\eeq
where $Q_{HI}=1-Q_{HII}$ is the neutral hydrogen filling factor
($Q_{HeI} = 1-Q_{HeII}-Q_{HeIII}$) and $\tau^1_{\nu,N_j}$ corresponds
to discrete clouds while $\tau^1_{\nu,IGM_j}$ describes the IGM
contribution. The typical distance $l_g(x)$ between galactic sources
characterized by their parameter $x$, density contrast $\Delta$ and
radius $R$, is given by their number density, see (\ref{etah}),
\beq
l_g(x) \sim  R \; (1+\Delta)^{1/3} \; \left[ x^2 H(x) \right]^{-1/3}
\eeq
where we did not take into account correlations. Since only a fraction
$\lambda_Q \leq 1$ of galaxies host quasars we have for the mean distance between bubbles ionized by quasars:
\beq
l_Q(x) = \left( \lambda_Q  \mbox{Min} \left[ 1 , \mbox{Max} \left( \frac{t_Q}{t_H} , \frac{t_{rec}}{t_H} \right) \right] \right)^{-1/3} \; l_g(x)   \label{distQ}
\eeq
as in (\ref{etabubbleQ}). Here $t_{rec}$ is the recombination time within the ionized bubbles, see Fig.\ref{figtionrecO03}. Since at low redshift $t_{rec} \sim t_H$ we usually have $l_Q \sim \lambda_Q^{-1/3} l_g$. Next, we define an effective opacity $\tau_{eff}$ over the region of size $l$ and volume $V$ by:
\beq
e^{-\tau_{eff}} = \int_0^{l} \; \frac{d^3r}{V} \; e^{-\tau(r)}
\label{taueff}
\eeq
where $\tau(r)$ is given by (\ref{tauQHI}). Then we use for the opacities which enter the source terms (\ref{Snus}) and (\ref{SnuQ}) a simple prescription which recovers the asymptotic regimes $\tau_{eff} \rightarrow 0$ and $\tau_{eff} \rightarrow \infty$ of (\ref{taueff}):
\beq
\tau_{s,Q}(x) = \ln \left( 1 + \frac{3}{4} \tau_{\nu}(l_{g,Q}) + \frac{1}{6}
\tau_{\nu}(l_{g,Q})^3 \right)    \label{tausQ}
\eeq

\section{Numerical results: open universe}
\label{Open universe}

We can now use the model we described in the previous sections to
obtain the reionization history of the universe, as well as many other
properties such as the  population of quasars, galaxies or Lyman-$\alpha$
clouds, for various cosmologies. We shall first consider the case of an
open universe $\Omega_0=0.3$, $\Lambda=0$, with a CDM power-spectrum (Davis et
al.1985), normalized to $\sigma_8=0.77$. We choose a baryonic density
parameter $\Omega_b=0.03$ and $H_0=60$ km/s/Mpc. We use the scaling
function $h(x)$ obtained by Bouchet et al.(1991) as explained in VS II. Our model is consistent with the studies presented in VS II and Valageas et
al.(1999a), so that those papers are part of the same unified model
and describe in more details our predictions for galaxies and
Lyman-$\alpha$ clouds at $z \leq 5$.

\subsection{Quasar luminosity function}
\label{Quasar luminosity function}

\begin{figure}[htb]

\begin{picture}(230,430)

\epsfxsize=26 cm
\epsfysize=18 cm
\put(-28,-50){\epsfbox{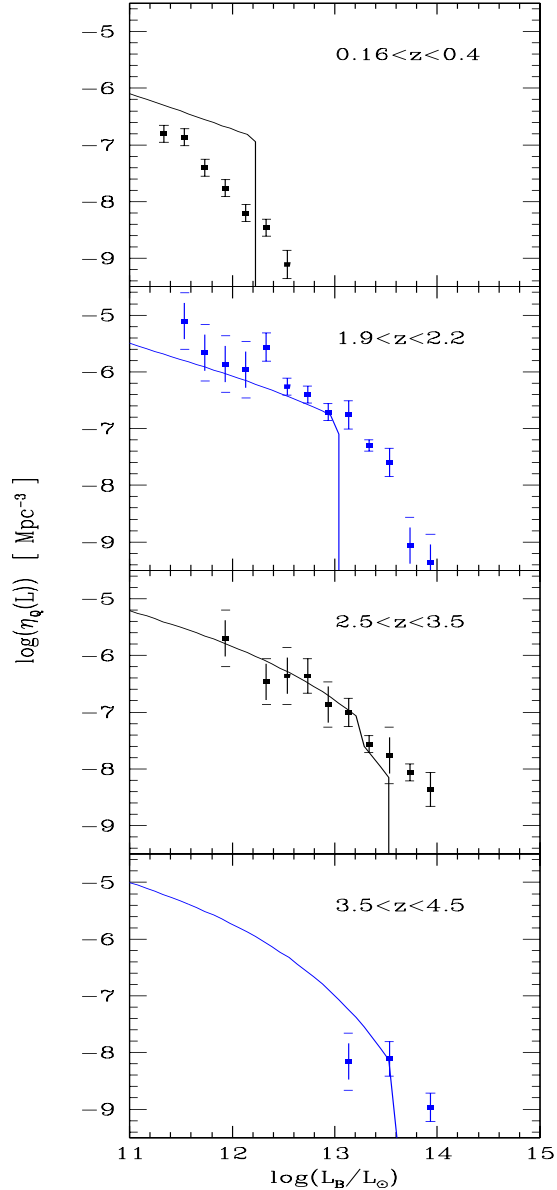}}

\end{picture}

\caption{The evolution with redshift of the B-band quasar luminosity 
function in comoving Mpc$^{-3}$. The data points are from Pei (1995).}
\label{figquasO03}

\end{figure}

Although our model was already checked in previous studies for
galaxies and Lyman-$\alpha$ clouds as explained above, our description
for quasars was not compared to observations in great detail
(although a first check was performed in VS II). Thus, we first
compare in this section our predictions for the quasar luminosity
function to observational data, as shown in Fig.\ref{figquasO03}.

We can see that our model is consistent with observations. At low redshifts the number of quasars we predict does not decline
as fast as the data, however we get a significant decrease which is
already an improvement over the results of Efstathiou \& Rees (1988)
for instance. We can note that Haehnelt \& Rees (1993) managed to
obtain a good fit to the observed decline at low $z$ but they had to
introduce an ad-hoc redshift and circular velocity dependence for the
black hole formation efficiency. Since our model appears to work
reasonably well we
prefer not to introduce additional parameters. Moreover, as we noticed
earlier our ratio (black hole mass)/(stellar mass) is consistent with
observations ($F=0.006$) while the quasar life-time we use $t_Q=0.44
\; 10^8$ yrs agrees with theoretical expectations. The high-luminosity
cutoff, which appears at $z<3.5$, comes from the fact that in our
model very massive and bright galaxies have consumed most of their
gas. Thus, the maximum quasar luminosity starts decreasing with time
at low $z$ because of fuel exhaustion. We  note that Haiman \& Loeb
(1998) obtained similar results at $z>2$ although they used a very
small time-scale $t_Q \sim 6.6 \; 10^5$ yrs (in our case this problem
is partly solved by the introduction of the parameter $\lambda_Q < 1$
which states that only a small fraction of galaxies actually host a
black hole). However, they note that the number density of bright
quasars they get increases until $z=0$.

\begin{figure}[htb]

\centerline {\epsfxsize=8 cm \epsfysize=5.5 cm \epsfbox{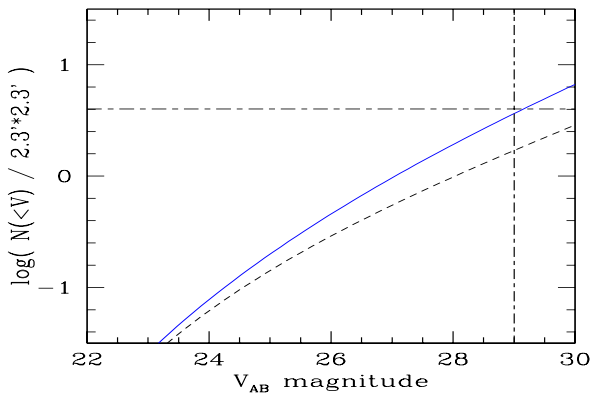}}

\caption{The quasar cumulative V-band number counts. The dashed line shows 
the counts of quasars with magnitude brighter than $V$ located at
the redshifts $3.5 < z < 4.5$ while the solid line corresponds to $3.5
< z < z_{ri}$.}
\label{figcountO03}

\end{figure}

In a recent paper, Haiman et al.(1998) point out that the lack of
quasar detection down to  magnitude $V=29$ in the HDF strongly  constrains
the models of quasar formation, which tend to predict more
than 4 objects (which is still marginally consistent). In particular,
they find that one needs to introduce a lower cutoff for the possible
mass of quasars (shallow potential wells with a circular velocity
lower than 50 km/s are not allowed to form back holes) or a
mass-dependent black-hole formation efficiency. We show in
Fig.\ref{figcountO03} the predictions of our model.

The solid line shows the number $N(<V)$ of quasars with a magnitude
lower than $V$ located at a redshift larger than $3.5$ up to the
reionization redshift $z_{ri}=6.8$. The large opacity beyond this
redshift prevents detection of higher $z$ objects. We also take into
account the opacity due to Lyman-$\alpha$ clouds at lower $z$. We can
see that our model is marginally consistent with the constraints from
the HDF since it predicts $4$ detections up to $V=29$. We note
that our model automatically includes  photoionization feedback
(threshold $T_{cool}$) and a virial temperature dependence in the
relation (black hole mass) - (dark matter halo mass). However, the
``cooling temperature'' $T_{cool}$ is too low to have a significant
effect on the number counts. Of course, we  see that at bright
magnitudes most of the counts come from low-redshift quasars
($z<4.5$). Thus, the QSO number counts strongly constrain our model
since in order to obtain a reionization history consistent with
observations (namely the HI and HeII Gunn-Peterson tests and the
low-redshift amplitude of the UV background radiation field) we need a
relatively large quasar multiplicity function. However, one might
weaken these constraints by using an ad-hoc QSO luminosity function
with many faint objects ($L_B < 5 \; 10^9 \; L_{\odot}$).

\subsection{Reheating of the IGM}

\begin{figure}[htb]

\centerline {\epsfxsize=8 cm \epsfysize=5.5 cm \epsfbox{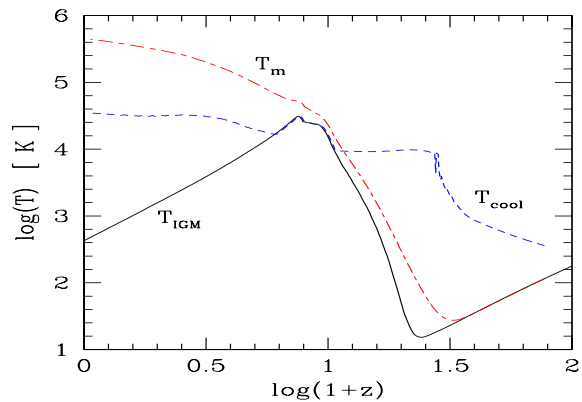}}

\caption{The redshift evolution of the IGM temperature $T_{IGM}$ 
(solid curve). We also show the virial temperature $T_{cool}$ of the
smallest virialized halos which can cool at redshift $z$ (dashed
curve) while $T_m$ is a mass-averaged temperature (dot-dashed curve).}
\label{figTO03}

\end{figure}

As we explained previously the radiation emitted by galaxies and
quasars will reheat and reionize the universe, following (\ref{TIGM})
and (\ref{Jnu}). We start our calculations at $z_i=200$ with the
initial conditions used by Abel et al.(1998), see also Peebles
(1993). In particular:
\[
T_{IGM}(z_i) = \mbox{Min} \left[\; 135 \; \left( \frac{1+z_i}{100} \right)^2 
\; \mbox{K} \; , \; 2.73 \; (1+z_i) \; \mbox{K} \right]
\]
\[
\frac{n_{HII}}{n_H} = 2.4 \; 10^{-4} \; \Omega_0^{1/2} \; 
\frac{0.05}{h \Omega_b}
\]
\[
\frac{n_{H_2}}{n_H} = 2 \; 10^{-20} \; \frac{(1-Y) \Omega_0^{3/2}}{h \Omega_b}
\; (1+z_i)^{5.1}
\]
and we use a helium mass fraction $Y=0.26$. We present in
Fig.\ref{figTO03} the redshift evolution of the IGM temperature
$T_{IGM}$, as well as the temperature $T_{cool}$ which defines the
smallest virialized objects which can cool at redshift $z$, see
(\ref{tcool}).

At high $z$ the IGM temperature decreases with time due to the
adiabatic expansion of the universe. Next, for $z < 24$ ($\log(1+z)<1.4$) the medium
starts being slowly reheated by the radiative output of stars and
quasars until it reaches at $z \simeq 9$ a maximum temperature
$T_{max} \sim 3 \; 10^4$ K where collisional excitation cooling is so
efficient that the IGM temperature cannot increase significantly any
more. As we shall see later, this phase occurs {\it before} the medium
is reionized, as was also noticed by Gnedin \& Ostriker (1997) using a
numerical simulation. There is a small increase at $z \sim 7$ ($\log(1+z) \sim 0.9$) when the
universe is fully reionized and the UV background radiation shows a
sharp rise. However, because cooling is very efficient, the dramatic
increase in $J_{\nu}$ only leads to a small change in
$T_{IGM}$. Eventually at low redshifts the temperature starts
decreasing again due to the expansion of the universe as the heating
time-scale becomes larger than the Hubble time $t_H$. The temperature
$T_{cool}$ which defines the smallest objects which can cool at
redshift $z$ increases with time because the decline of the number
density of the various species, due to the expansion of the universe,
makes cooling less and less efficient. Indeed, the cooling rate (in
erg cm$^{-3}$ s$^{-1}$) associated with a given process involving the
species $i$ and $j$ can usually be written as $k_{ij}(T) n_i n_j$,
which leads to a cooling time-scale:
\beq
t_{cool,ij} = \frac{3/2 \; n_b k T}{k_{ij}(T) n_i n_j} \propto n^{-1} \sim
(1+z)^{-3/2} \; t_H
\eeq
where we have neglected the temperature dependence. Thus, the ratio
(cooling time)/(Hubble time) increases as time goes on (at fixed $T$
and abundance fractions). Since halos with virial temperature
$T_{cool}$ must satisfy $t_{cool} \sim t_H$, see (\ref{tcool}),
$T_{cool}$ has to get higher with time to increase the rate $k_{ij}$
(which usually contains factors of the form $\exp(-T_{ij}/T)$). The
sudden increase of $T_{cool}$ at $z \sim 30$ ($\log(1+z) \sim 1.5$) is due to the decline of
the fraction of molecular hydrogen which starts being destroyed by the
radiation emitted by stars and quasars. As a consequence the main
cooling process becomes collisional excitation cooling instead of
molecular cooling. Since the former is only active at high
temperatures (the coefficient rate $k(T)$ contains a term
$\exp(-118348 K/T)$ instead of $ \exp(-512 K/T)$ for molecular
cooling) the cooling temperature $T_{cool}$ has to increase up to
$T_{cool} \sim 10^4$ K. By definition $T_{cool}$ is larger than the
IGM temperature and usually much higher as can be seen in
Fig.\ref{figTO03}. However, at $z \sim 9$ when $T_{IGM} \sim 10^4$ K
is quite high due to reheating by the background radiation field we
have $T_{cool}=T_{IGM}$ since the IGM temperature is large enough to
allow for efficient cooling. Then all virialized bound objects, with
$T>T_{IGM}$, form baryonic clumps which can cool. The temperature
$T_m$ represents a mass-averaged temperature: the matter within the
IGM is associated to $T_{IGM}$ while virialized objects (hence with
$T>T_{IGM}$) are characterized by a temperature defined as Min$(T,10^6
\mbox{K})$. Since $T_m$ does not enter any of our calculations used to
obtain the redshift evolution of the universe this crude definition is
sufficient for our purpose which is merely to illustrate the
difference between volume ($T_{IGM}$) and mass averages. As can be
seen from Fig.\ref{figTO03} we always have $T_{m} \geq T_{IGM}$ as it
should be. At large redshifts $T_{m} \simeq T_{IGM}$ since most of the
matter is within the uniform IGM component, whereas at low redshifts $z
< 5$ ($\log(1+z) < 0.8$) the IGM temperature declines as we explained previously while
$T_m$ remains large since most of the matter is now embedded within
collapsed objects where shock heating is important (and they do not
experience adiabatic cooling due to the expansion).

\begin{figure}[htb]

\centerline{\epsfxsize=8 cm \epsfysize=11.5 cm \epsfbox{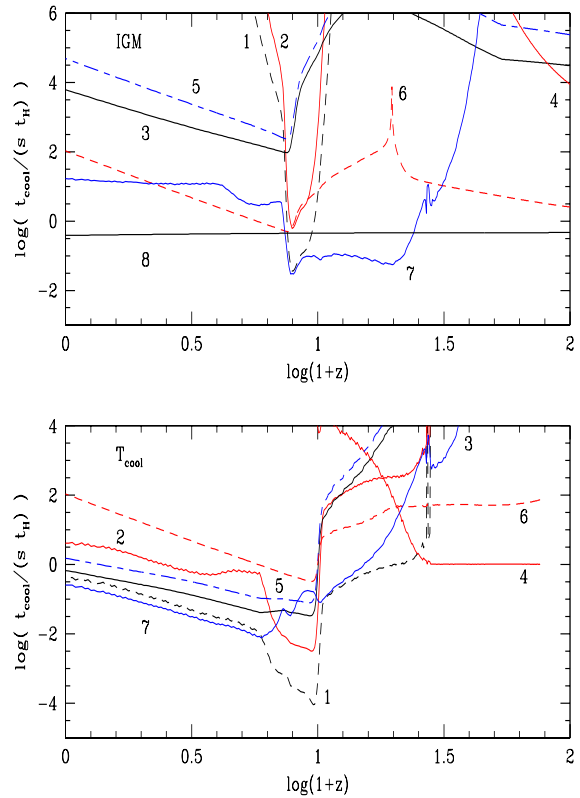}}

\caption{The cooling and heating times associated with the various 
relevant processes in units of $s \; t_H(z)$ for the IGM (upper
figure) and the halos defined by $T_{cool}$ (lower figure). The labels
are as follows: 1) collisional excitation, 2) collisional ionization,
3) recombination, 4) molecular hydrogen, 5) bremsstrahlung, 6)
Compton, 7) photoionization heating and 8) cooling due to expansion
(only for the IGM, see text).}
\label{figtcoolO03}

\end{figure}

We show in Fig.\ref{figtcoolO03} the cooling and heating times
associated with various processes for the IGM as well as for the
smallest halos $M_{cool}$ which can cool at $z$.

We can see in the upper panel that for large and small redshifts, $z >
24$ ($\log(1+z) > 1.4$) and $z < 5$ ($\log(1+z) < 0.8$), all time-scales associated with the IGM are larger
than the Hubble-time which means that the IGM temperature declines due
to the adiabatic cooling entailed by the expansion of the
universe. However, at intermediate redshifts $z \sim 18$ ($\log(1+z) \sim 1.3$) the smallest
time-scale corresponds to heating by the background radiation
($t_{heat}$) which means that $T_{IGM}$ increases during this
period. Next, at $z \sim 9$, the IGM temperature becomes large enough
to activate collisional excitation cooling so as to reach a temporary
equilibrium where $t_{cool} \simeq t_{heat}$ while $T_{IGM}$ remains
constant. Then, as we shall see later the universe gets suddenly
reionized at $z_{ri} = 6.8$ ($\log(1+z)=0.9$). This means that $t_{heat}$ increases
sharply as $n_{HI}$ declines (as well as $n_{HeI}$ and $n_{HeII}$) as
can be seen from (\ref{theat}). The cooling time due to collisional
excitation follows this rise as the medium remains in
quasi-equilibrium while the temperature declines slightly (the strong
temperature-dependent factors like $\exp(-118348 K/T)$ in $t_{cool}$
ensure it immediately adjusts to $t_{heat}$, moreover these cooling
rates are also proportional to $n_{HI}$, $n_{HeI}$ and $n_{HeII}$)
until both heating and cooling time-scales become larger than the
Hubble time. Then this quasi-equilibrium regime stops as the medium
merely cools because of the expansion of the universe. These various
phases, which appear very clearly in Fig.\ref{figtcoolO03}, explain the
behaviour of $T_{IGM}$ shown in Fig.\ref{figTO03} which we described
earlier. The peak at $z \simeq 19$ ($\log(1+z) \simeq 1.3$) of $t_{Compton}$ in the upper panel
(curve 6) corresponds to the time when its sign changes (hence
$t_{Compton}^{-1}=0$). At higher redshifts $T_{IGM}$ is lower than the
CMB temperature (due to adiabatic cooling by the expansion of the
universe), so that the gas is heated by the CMB photons while at lower
$z$ the IGM temperature is larger than $T_{CMB}$ (due to reheating) so
that the gas is cooled by the interaction with the CMB radiation.

The lower panel shows the cooling and heating times associated with
the halos $T_{cool}$. As we have already explained we can see that at
high redshifts the main cooling process is molecular hydrogen
cooling. Note however that for the IGM this process is always
irrelevant. This difference comes from the fact that the larger
density and temperature of these virialized halos allow them to form
more molecular hydrogen than is present in the IGM, so that molecular
cooling becomes efficient. This was also described in detail in
Tegmark et al.(1997) for instance. Of course at these redshifts we
have $t_{cool,mol} = s
\; t_H$ by definition of $T_{cool}$. Then, at $z < 27$ ($\log(1+z) < 1.4$) as molecular
hydrogen starts being destroyed by the background radiation the main
cooling process becomes collisional excitation. Note that the
corresponding cooling time gets smaller than $s \; t_H$ because the
medium is also heated by the radiation so that the actual cooling time
results from a slight imbalance between cooling and heating
processes. The sharp decrease of the various time-scales at $z \simeq 9$ corresponds to a sudden increase of $T_{cool}$ due to the rise of $T_{IGM}$ (which influences
the cooling halos since $T_{cool} \geq T_{IGM}$) also seen in
Fig.\ref{figTO03} and in the upper panel of
Fig.\ref{figtcoolO03}. Around $z \sim 8$ ($\log(1+z) \sim 0.9$) we have $t_{cool} \ll
t_{heat}$ and $t_{cool} \ll t_H$ so that all virialized halos above
$T_{IGM}$ can cool ($T_{cool}=T_{IGM}$). The feature at $z \sim 5.3$ ($\log(1+z) \sim 0.8$) is due to reionization.

\begin{figure}[htb]

\centerline {\epsfxsize=8 cm \epsfysize=5.5 cm \epsfbox{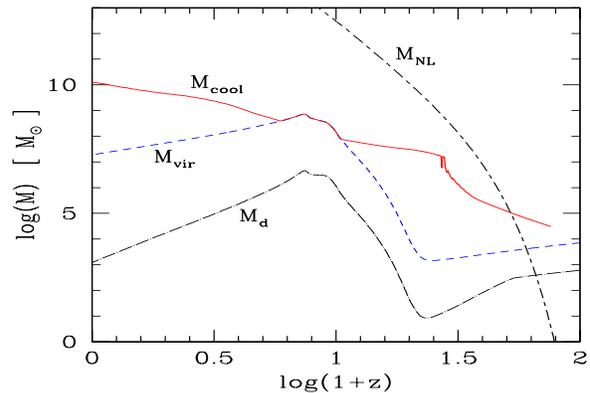}}

\caption{The redshift evolution of the characteristic masses $M_d$, 
$M_{vir}$, $M_{cool}$ and $M_{NL}$ in $M_{\odot}$.}
\label{figMO03}

\end{figure}

Finally, we present in Fig.\ref{figMO03} the characteristic masses we
encounter. The mass $M_d$ is obtained from (\ref{Rd}):
\beq
M_d = (1+\Delta)_{IGM} \; \rhoa \frac{4 \pi}{3} R_d^3
\label{Md}
\eeq
while $M_{cool}$ corresponds to the halos which can cool at redshift $z$, as
we have already explained. The mass $M_{vir}$ which follows closely
the behaviour of $M_d$ describes the smallest virialized objects. It
differs from $M_d$ because the density contrast is now $\Delta_c$
instead of $\Delta_{IGM}$. The mass $M_{NL}$ corresponds to the first
non-linear scale defined by $\xia = 1$. Note that after reionization $M_d \sim 3 \; 10^6 M_{\odot}$ while the usual Jeans mass would be $M_J \sim 10^8 M_{\odot}$. This is mainly due to the low density $(1+\Delta)_{IGM}$ of the IGM, see (\ref{Md}) and Fig.\ref{figclumpO03} below.

By definition we have $M_d < M_{vir} \leq M_{cool}$. At $z \sim 9$ we
have $M_{cool} = M_{vir}$ since $T_{cool}=T_{IGM}$ as we noticed
earlier in Fig.\ref{figTO03}. We also note that at large $z$ our
calculation is not entirely correct since our multiplicity functions
are valid in the non-linear regime, for masses $M \ll
M_{NL}$. However, at these early times the universe is nearly exactly
uniform (by definition !) so that this is not a very serious
problem. We can see that the first cooled objects which form in
significant numbers are halos of dark-matter mass $M \sim 10^5
M_{\odot}$ which appear at $z \sim 49$ ($\log(1+z) \sim 1.7$), when $M_{cool}$ becomes
smaller than $M_{NL}$. However, they only influence the IGM after $z <
24$ ($\log(1+z) < 1.4$) when reheating begins.

\subsection{Reionization of the IGM}

\begin{figure}[htb]

\centerline {\epsfxsize=8 cm \epsfysize=11.5 cm \epsfbox{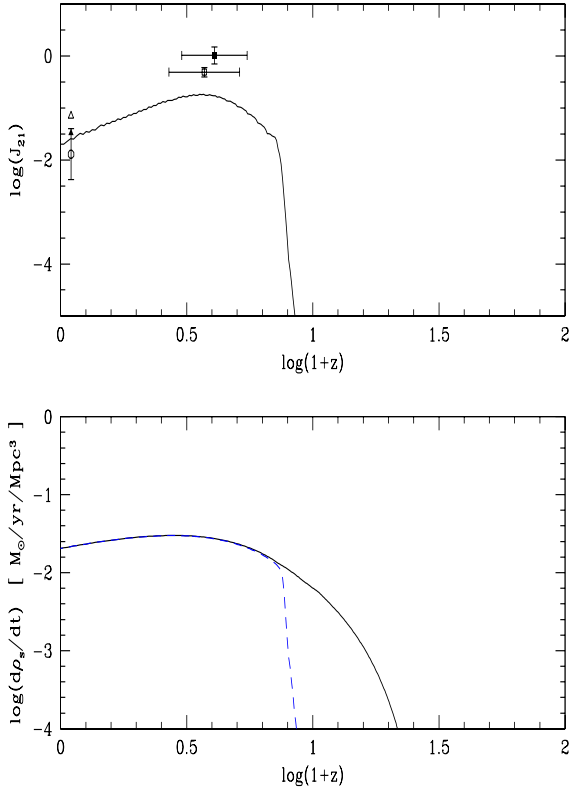}}

\caption{The redshift evolution of the UV flux $J_{21}$ (upper panel) and 
of the comoving star formation rate $d\rho_s/dt$ (lower panel). The
data points are from Giallongo et al.(1996) (square), Cooke et
al.(1997) (filled square), Vogel et al.(1995) (triangle, upper limit),
Donahue et al.(1995) (filled triangle, upper limit) and Kulkarni \&
Fall(1993) (circle). The dashed line in the lower panel shows the
effect of the absorption of high energy photons by the neutral
hydrogen present in the IGM and in Lyman-$\alpha$ clouds.}
\label{figJO03}

\end{figure}

After the radiation emitted by quasars and stars reheats the universe,
as described in the previous section, it will eventually reionize the
IGM. We present in Fig.\ref{figJO03} the evolution with redshift of the
background radiation field and of the comoving stellar formation
rate. Within the framework of our model the latter is a good measure
of the radiative output from galaxies, see (\ref{dEdnudts}), as well
as from quasars, see (\ref{LQ}), since we note that the quasar mass
happens to be roughly proportional to the stellar mass.

The upper panel of Fig.\ref{figJO03} shows the UV flux $J_{21}$ in
units of $10^{-21}$ erg cm$^{-2}$ s$^{-1}$ Hz$^{-1}$ sr$^{-1}$ defined
by (\ref{J21}). We can see that the UV flux rises very sharply at $z
\simeq 6.8$ ($\log(1+z) \simeq 0.9$) which corresponds to the reionization redshift $z_{ri}$ when
the universe suddenly becomes optically thin, so that the radiation
emitted by stars and quasars at large frequencies is no longer
absorbed and contributes directly to $J_{\nu}$. This appears clearly
from the lower panel. Here the solid line shows the comoving star
formation rate, obtained from (\ref{SFRav}), while the dashed line
shows the same quantity multiplied by a luminosity-weighted opacity
factor $\exp(-\tau_L)$ which describes the opacity due to the IGM and
Lyman-$\alpha$ clouds (see below (\ref{tauL}), (\ref{tauLtot}) and
Fig.\ref{figtauO03}). Thus, we can see that while the star formation
rate evolves rather slowly with $z$ the absorption term varies sharply
around $z_{ri}$. Hence the universe is suddenly reionized at $z_{ri}$
(when the ionized bubbles overlap: $Q_{HII}=1$) on a time-scale very
short as compared to the Hubble time because of the strongly
non-linear effect of the opacity. We note that for $z<1$ our
star-formation model is somewhat simplified, as we explained earlier,
because we defined all galaxies by a constant density contrast
$\Delta_c$ while cooling constraints should be taken into account, as
in VS II, and we also used approximations for the stellar content of
galaxies which are not strictly valid for all galactic halos at these
low redshifts. The reader is referred to VS II for a more precise
description of the low $z$ behaviour. However, our present treatment
is sufficient for our purposes and still provides a reasonable
approximation at $z < 1$.

\begin{figure}[htb]

\centerline {\epsfxsize=8 cm \epsfysize=5.5 cm \epsfbox{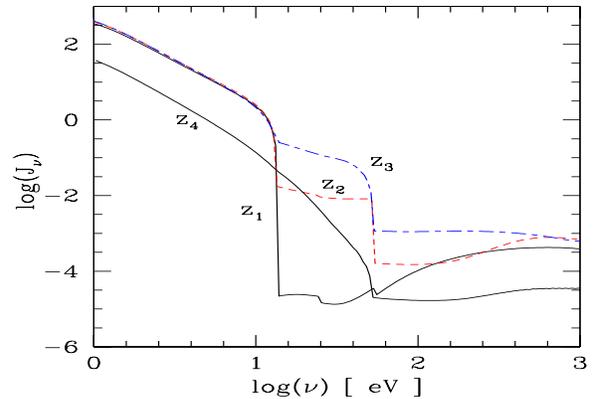}}

\caption{The background radiation spectrum $J_{\nu}$ (in
units of $10^{-21}$ erg cm$^{-2}$ s$^{-1}$ Hz$^{-1}$ sr$^{-1}$) at the
redshifts $z_1=7.3$ (solid line, prior to reionization), $z_2=6.4$
(lower dashed line, after reionization), $z_3=3$ (upper dashed
line) and $z_4=0$ (solid line).}
\label{figJnuO03}

\end{figure}

We display in Fig.\ref{figJnuO03} the background radiation spectrum
$J_{\nu}$ at four redshifts: $z_1=7.3$ (before reionization),
$z_2=6.4$ (after reionization), $z_3=3$ and $z_4=0$.

We see that the ionization edges corresponding to HI, HeI and HeII can
be clearly seen at high $z$ before the universe is reionized. Then the
background radiation is very strongly suppressed for $\nu > 13.6$ eV
due to HI and HeI absorption. Of course at very large frequencies $\nu
\ga 1$ keV where the cross-section gets small we recover the slope
$J_{\nu} \propto \nu^{-\alpha}$ of the radiation emitted by
quasars. At low redshifts $z < z_{ri}$ after reionization the drop
corresponding to HeI disappears as HeI is fully ionized and its number
density gets extremely small, as we shall see below in
Fig.\ref{figchemO03}. However, even at $z \sim 3$ the ionization edges
due to HI and HeII are clearly apparent and $J_{\nu}$ is still
significantly different from a simple power-law. At low redshifts $z
\sim 0$ the background radiation is much smoother since its main
contribution comes from radiation emitted while the universe was
ionized and optically thin. However, its intensity is smaller than at
$z \sim 4$ because the quasar luminosity function drops at low $z$,
see Fig.\ref{figquasO03}, while the universe keeps expanding.

\begin{figure}[htb]

\centerline {\epsfxsize=8 cm \epsfysize=5.5 cm \epsfbox{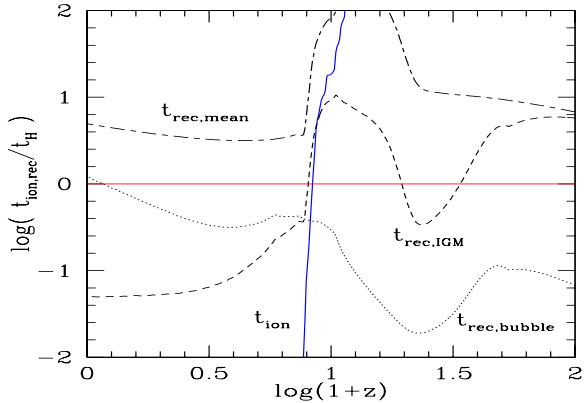}}

\caption{The redshift evolution of the ionization and recombination times 
$t_{ion}$ (solid line) and $t_{rec,IGM}$ (dashed line) of the IGM, divided
by the Hubble time $t_H$. The horizontal solid line only shows $t_H$
for reference. The recombination times $t_{rec,mean}$ (uniform medium) and $t_{rec,bubble}$ (ionized bubbles) are defined in the main text.}
\label{figtionrecO03}

\end{figure}

We show in Fig.\ref{figtionrecO03} the redshift evolution of the
ionization and recombination times $t_{ion}$ and $t_{rec}$ of the IGM,
divided by $t_H$.

More precisely, the ionization time $t_{ion}$ is defined by:
\beq
t_{ion}^{-1} = \int 4 \pi \; \frac{J_{\nu}}{h \nu} \; \sigma_{HI}(\nu)
\; d\nu     \label{tion}
\eeq
while the recombination time within the IGM is:
\beq
t_{rec,IGM}^{-1} = \alpha(T_{IGM}) \; C_n \; \nbar_{e-}    \label{trec}
\eeq
where $\alpha$ is the recombination rate, $C_n$ the clumping
factor and $\nbar_{e-}$ the mean electron number density, from (\ref{Cn}) and (\ref{rhon}). We also display for reference the recombination time which would correspond to a uniform medium with the mean density of the universe:
\beq
t_{rec,mean} = (1+\Delta)_n \; C_n \; t_{rec,IGM}
\eeq
Finally, we show the recombination time within ionized bubbles (where all hydrogen atoms are ionized):
\beq
t_{rec,bubble}^{-1} = \alpha \left(3 \; 10^4 \mbox{K} \right) \; C_n \; \nbar_H   \label{trecbubble}
\eeq
The recombination time grows with time at high $z$ as the mean density
decreases with the expansion, although this is somewhat balanced by
the increase of the clumping factor $C_n$ (see below
Fig.\ref{figclumpO03}). In particular, the decrease of $t_{rec,IGM}$ around $z  \sim 30$ ($\log(1+z) \sim 1.5$) is due to the growth of $C_n$. The sharp drop at $z \sim 7$ ($\log(1+z) \sim 0.9$) is due to
reionization which suddenly increases the number density of free
electrons. After reionization the recombination time characteristic of the IGM keeps decreasing (while $t_{rec,mean}$ increases slightly since the density declines) because of the growth of the clumping factor $C_n$, see Fig.\ref{figclumpO03}, which overides the decline of the mean universe density. The recombination time within ionized bubbles $t_{rec,bubble}$ follows the change of the mean universe density and of the clumping factor $C_n$. At large $z$ it is much smaller than the mean IGM recombination time since the IGM is close to neutral. At low $z$ it becomes larger than $t_{rec,IGM}$ since the IGM is suddenly reionized with a temperature $T_{IGM}$ which declines after reheating and gets lower than $3 \; 10^4$ K, see Fig.\ref{figTO03}. The ionization time is very large at high $z$ since the UV
background radiation is small. Then it decreases very sharply at
$z_{ri}$ when the universe is reionized and the background radiation
suddenly grows as the medium becomes optically thin, as seen in
Fig.\ref{figJO03}. The reionization redshift corresponds to the time
when $t_{ion}$ becomes smaller than $t_H$, somewhat after it gets
smaller than $t_{rec,IGM}$. Thus $t_{rec,IGM}$ does not play a decisive role
since it is never the smallest time-scale around reionization.

Finally, we show in Fig.\ref{figchemO03} how the chemistry of the IGM
evolves with time as the temperature $T_{IGM}$ and the UV flux
$J_{21}$ vary with $z$.

\begin{figure}[htb]

\centerline {\epsfxsize=8 cm \epsfysize=11.5 cm \epsfbox{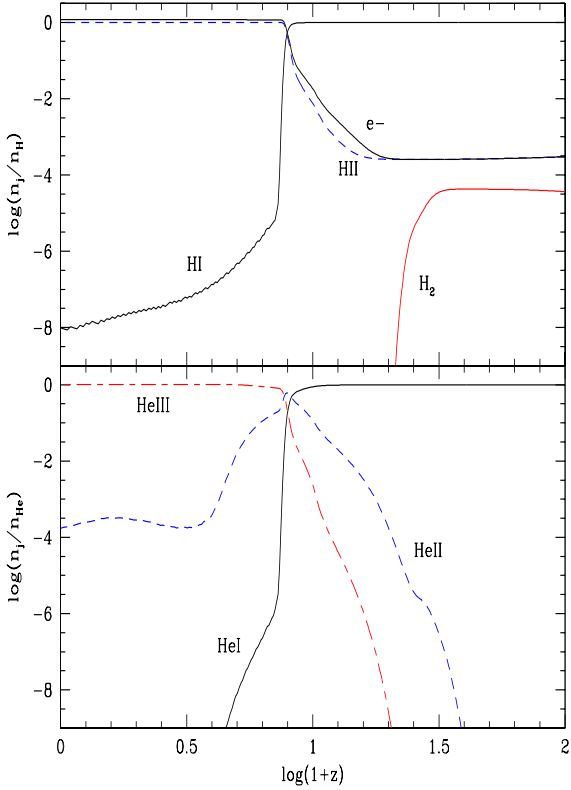}}

\caption{The redshift evolution of the chemistry of the IGM. The upper 
panel shows the ionization state of hydrogen, as well as the fraction
of molecular hydrogen and electrons. The lower panel presents the
ionization of helium.}
\label{figchemO03}

\end{figure}

We see very clearly in the upper panel the redshift of reionization
$z_{ri}=6.8$ ($\log(1+z) = 0.9$) when the fraction of neutral hydrogen $n_{HI}/n_{H}$
declines very sharply while the UV flux $J_{21}$ suddenly rises, as
was shown in Fig.\ref{figJO03}. We note that at low redshifts $z <
z_{ri}$ the neutral hydrogen fraction decreases more slowly down to
$n_{HI}/n_{H} \sim 10^{-8}$. The fractions of electrons and ionized
hydrogen start increasing earlier at $z \sim 19$ ($\log(1+z) \sim 1.3$) but of course they
remain small until $z_{ri}$. The abundance of molecular hydrogen
decreases sharply at a rather high redshift $z \sim 27$ ($\log(1+z) \sim 1.4$) due to the
background radiation, as we noticed on Fig.\ref{figtcoolO03}. The lower
panel shows that helium gets fully ionized simultaneously with
hydrogen. In particular, although there remains a small fraction of
HeII ($\sim 10^{-4}$) the abundance of HeI gets extremely small. We
note that at low redshifts $z < 2$ the fraction of HeII does not
evolve much (and even slightly increases) while the HI abundance keeps
declining. This is due to the fact that the radiation relevant for
helium ionization comes from quasars whose luminosity function drops
at low $z$ as shown in Fig.\ref{figquasO03} while an important
contribution to the hydrogen ionizing radiation is provided by stars
and the galaxy luminosity function declines more slowly with time at
low $z$, as seen in Fig.\ref{figJO03} or in VS II. We shall come back
to this point in Sect.\ref{Contributions of quasars and stars}.

\subsection{Opacities}

As we explained previously the radiation emitted by stars and quasars
at high frequencies ($\nu > 13.6$ eV) is absorbed by the IGM and
discrete clouds as it propagates into the IGM. This leads to the
extinction factors $\exp(-\tau_s(x))$ and $\exp(-\tau_Q(x))$ in the
evaluation of the source terms (\ref{Snus}) and (\ref{SnuQ}) for the
background radiation. We define here the ``luminosity averages''
$\tau^L_{IGM,NHI}$ for both continuous and discrete components by:
\beq
e^{ -\tau^L } = \left( \frac{d\rho_s}{dt} \right)^{-1}
\;  \int \eta_g(x) \; \frac{dx}{x} \; \left( \frac{dM_s}{dt} \right)(x)
\; e^{-\tau_s(x)}  \label{tauL}
\eeq
where $\tau_s(x)$ is the relevant opacity (from the IGM or clouds for
sources $x$, see (\ref{tausQ}) ) at the frequency 20 eV below the HeI
ionization threshold without taking into account the factors $Q_{HI}$
which describe ionized bubbles. The subscript $s$ refers to the fact
that we consider here the opacity which enters into the calculation of
the stellar radiative output. The quasar-related opacity mainly differs
through the factor $\lambda_Q$, see (\ref{distQ}). The weight
$dM_s/dt$ corresponds to a luminosity weight as we noticed earlier,
see (\ref{dEdnudts}), (\ref{LQ}) and (\ref{SFR}).

\begin{figure}[htb]

\centerline {\epsfxsize=8 cm \epsfysize=11.5 cm \epsfbox{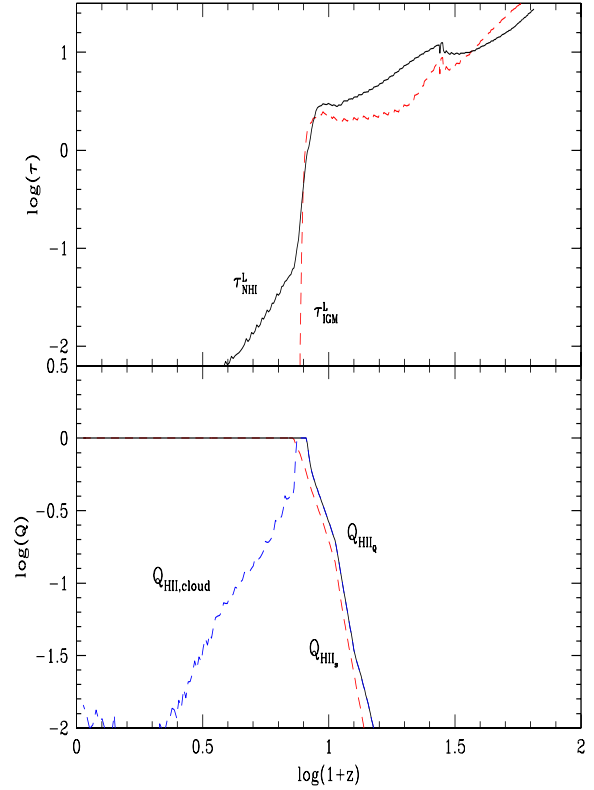}}

\caption{Upper panel: the redshift evolution of the opacity from the IGM 
(dashed line) and ``Lyman-$\alpha$ clouds'' (solid line) which enters
the absorption factors in the calculation of the radiative output from
stars and quasars. Lower panel: evolution of the filling factors
$Q_{HII_Q}$ (ionized bubbles around quasars, solid line), $Q_{HII_s}$
(around galaxies, upper dashed line) and $Q_{HII,cloud}$ (lower dashed
line).}
\label{figtauO03}

\end{figure}

We show in the upper panel of Fig.\ref{figtauO03} the redshift
evolution of the opacity from the IGM ($\tau^L_{IGM}$, dashed line)
and from ``Lyman-$\alpha$ clouds'' ($\tau^L_{NHI}$, solid line). We
can see that both contributions have roughly the same magnitude before
reionization, except at very high redshifts $z > 50$ ($\log(1+z) > 1.7$) when very few
structures exist as shown in Fig.\ref{figFracO03}. Prior to
reionization the opacity is large and the background radiation quite
small, as seen in Fig.\ref{figJO03}. At $z_{ri}$ the opacity suddenly
declines while $J_{21}$ rises sharply, due to the strong non-linear
coupling between $\tau$ and $J_{\nu}$, as we explained in the lower
panel of Fig.\ref{figJO03} where we presented the influence of the
total opacity $\tau_L$:
\beq
\tau_L = \tau^L_{IGM} + \tau^L_{NHI}
\label{tauLtot}
\eeq
At low redshifts $z < z_{ri}$ when the universe is reionized the
opacity due to discrete clouds becomes much larger than the IGM
contribution (although it is very small) because the density of
neutral hydrogen is now proportional to the square of the baryonic
density (in photoionized regions) and most of the matter is embedded
within collapsed objects.

The opacities $\tau^L$ were shown in the upper panel of
Fig.\ref{figtauO03} without the filling factors $Q_{HI}$ which enter
the actual evaluation of the source terms (\ref{Snus}) and
(\ref{SnuQ}), see (\ref{tauQHI}). The filling factors $Q_{HII}$,
describing the volume fraction occupied by ionized bubbles, are shown
in the lower panel of Fig.\ref{figtauO03}. We can see that $Q_{HII_s}$
and $Q_{HII_Q}$ increase with time as structures form and emit
radiation while the IGM density declines. When these ionized bubbles
overlap ($Q_{HII}=1$) the universe is reionized. At low redshifts the
coefficient $Q_{HII,cloud}$ declines because the background radiation
is large while the quasar number density drops (indeed $Q_{HII,cloud}$
measures the volume occupied by the ``spheres of influence'' of
quasars).

\begin{figure}[htb]

\centerline {\epsfxsize=8 cm \epsfysize=5.5 cm \epsfbox{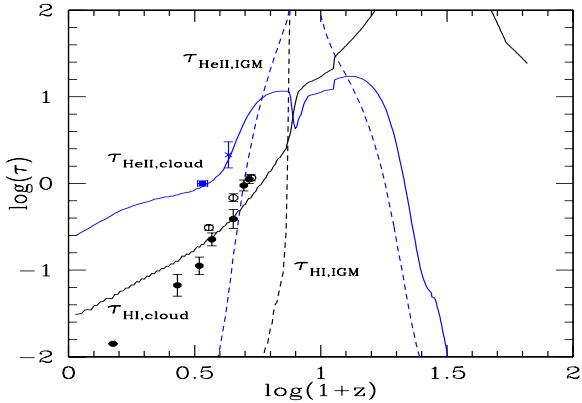}}

\caption{The redshift evolution of the average opacities $\tau_{HI}$ and 
$\tau_{HeII}$ along a random line of sight produced by
``Lyman-$\alpha$ clouds''. The dashed lines show the opacities from
the uniform IGM. The data points are from Press et al.(1993)
(circles), Zuo \& Lu (1993) (filled circles) for hydrogen, and from
Davidsen et al.(1996) (filled rectangle) and Hogan et al.(1997)
(cross) for helium.}
\label{figtauLymO03}

\end{figure}

Next, we can evaluate the mean opacities $\tau_{HI}(z)$ and
$\tau_{HeII}(z)$ seen on a random line of sight from $z=0$ to a quasar
located at redshift $z$. We present in Fig.\ref{figtauLymO03} the
contributions from both the uniform IGM component and the discrete
Lyman-$\alpha$ clouds.

At high redshifts prior to reionization the main contribution to the
opacity is provided by the IGM which contains most of the
matter. However, at low $z$ when the IGM is ionized most of the
absorption comes from the Lyman-$\alpha$ clouds. At large $z$ the HeII
opacity is very small because most of the helium is in its neutral
form HeI. We refer the reader to Valageas et al.(1999a) for a much
more detailed description of the properties of the Lyman-$\alpha$
clouds at low $z$. We can see that our predictions show a reasonable
agreement with observations for the hydrogen opacity. At low redshifts
$z < 1$ the influence of star-formation which consumes and may eject
some of the gas (which we did not take into account here) could
explain the relatively high opacity we obtain. The helium opacity we
find is also close to observations. This is due to the fact that the UV
radiation spectrum shows strong ionization edges, even at relatively
low redshifts $z \sim 3$ ($\log(1+z) \sim 0.6$), see Fig.\ref{figJnuO03}. Hence the ratio
$N_{HeII}/N_{HI}$ is rather large, see Fig.\ref{figchemO03}, which
explains why we get a better agreement than Zheng et al.(1998) for
instance (see also Valageas et al.1999a). In particular, we have $n_{HeII}/n_{HeIII} \gg n_{HI}/n_{HII}$ since $n_{HeII}/n_{HeIII} \sim 10^{-4}$ while $n_{HI}/n_{HII} \sim 10^{-8}$ at low redshift. We note that the observed
HeII opacity strongly constrains the quasar contribution to the
reionization process since stellar radiation is small at high
frequencies (due to the near blackbody behaviour of stellar spectra). In
particular, it implies that one needs a population of faint QSOs
($M_B>-26.7$) in order to reionize helium but $z_{ri}$ should not be
too large so that there is still an appreciable density of HeII. In
other words, as we noticed above, the UV radiation field must still
display strong ionization edges, which means that it has not had enough
time to be smoothed out by the radiation emitted since $z_{ri}$ when
the medium is optically thin.

\subsection{Stellar properties}

Our model also allows us to obtain the fraction of matter within
virialized or cooled halos, as well as in stars.

\begin{figure}[htb]

\centerline {\epsfxsize=8 cm \epsfysize=5.5 cm \epsfbox{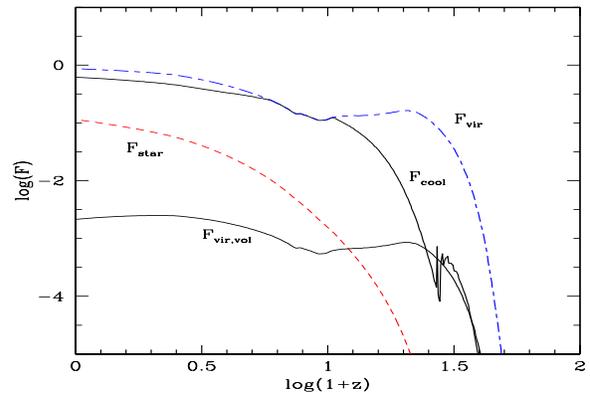}}

\caption{The redshift evolution of the fraction of matter 
enclosed within virialized halos ($F_{vir}$), cooled objects
($F_{cool}$) and stars ($F_{star}$). The lower solid line is the
volume fraction $F_{vir,vol}$ occupied by virialized objects.}
\label{figFracO03}

\end{figure}

We show in Fig.\ref{figFracO03} the fraction of matter within
virialized halos ($F_{vir}$, upper dashed line), cooled objects
($F_{cool}$, upper solid line) and stars ($F_{star}$, lower dashed
line). The first two quantities are simply obtained from
(\ref{muh}). Of course we have: $F_{star} \leq F_{cool} \leq
F_{vir}$. Around $z \sim 9$ we note that $F_{cool} = F_{vir}$ because
as we explained previously at this time all virialized objects (with
$T \geq T_{IGM}$) can cool efficiently ($T_{cool}=T_{IGM}$). The
fraction of matter within virialized halos increases very fast at high
redshifts $z \sim 49$ ($\log(1+z) \sim 1.6$) as $M_{vir}$ becomes smaller than $M_{NL}$, see
Fig.\ref{figMO03}, when dark matter structures form on scale
$R_d$. However, until $z \sim 15$ ($\log(1+z) \sim 1.2$) the mass within cooled halos remains
much smaller because cooling is not very efficient so that $T_{cool}
\gg T_{IGM}$, see Fig.\ref{figTO03}. At low redshifts $z < 5$ ($\log(1+z) < 0.8$) both
$F_{vir}$ and $F_{cool}$ get close to unity since most of the matter
is now embedded within collapsed and cooled halos (even though
$T_{cool}$ becomes again much larger than $T_{IGM}$: we are so far
within the non-linear regime that even $T_{cool}$ is small compared to
the characteristic virial temperature of the structures built on scale
$R_d$). Of course the mass within stars grows with time, closely
following $F_{cool}$. Note however that it is not strictly
proportional to $F_{cool}$ since an increasingly large fraction of the
gas within galaxies is consumed into stars. The fraction of volume
$F_{vir,vol}$ occupied by virialized objects always remains small as
it satisfies:
\beq
F_{vir,vol} = \frac{1}{1+\Delta_c} \; F_{vir} \leq
\frac{1}{1+\Delta_c}
\eeq

\begin{figure}[htb]

\centerline {\epsfxsize=8 cm \epsfysize=10 cm \epsfbox{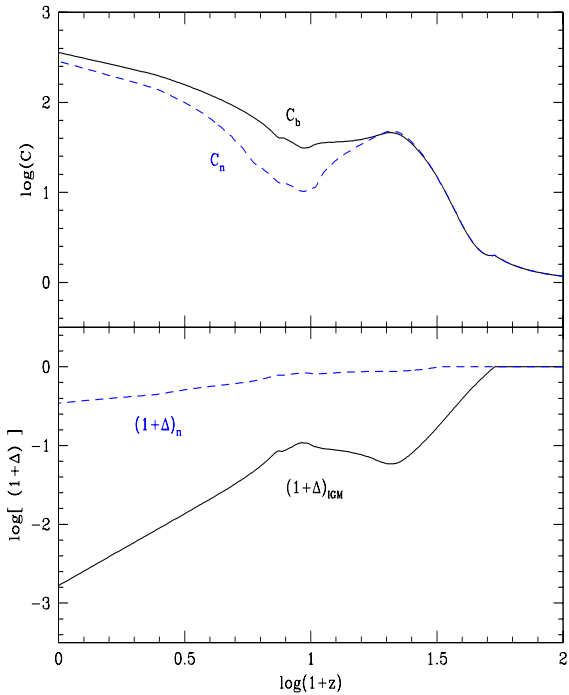}}

\caption{Upper panel: the redshift evolution of the clumping factors $C_b$ and $C_n$. Lower panel: the overdensities $(1+\Delta)_{IGM}(z)$ characteristic of
most of the volume of the universe at redshift $z$ at scale $R_d$ and $(1+\Delta)_n$.}
\label{figclumpO03}

\end{figure}

We show in the upper panel of Fig.\ref{figclumpO03} the clumping factor
$C_b$ defined by (\ref{Cb}). The expression (\ref{Cb}) shows clearly
that at large redshifts where there are very few collapsed baryonic
structures $C_b \simeq 1$ while at low $z$ when most of the baryonic
matter is within virialized halos (note this is always true for dark
matter on sufficiently small scales) we have $C_b \simeq \Delta_c$. We
can see in the figure that $C_b$ usually increases with time as the
hierarchical clustering process goes on. The temporary decrease at $z
\sim 15$ ($\log(1+z) \sim 1.2$), which also appears in the lower panel and in
Fig.\ref{figFracO03}, is due to the reheating of the universe, shown in
Fig.\ref{figTO03}, which increases the ``damping'' length $R_d$ and mass scale $M_d$, as
seen in Fig.\ref{figMO03}. As a consequence, small objects which were
previously well-defined entities suddenly see the IGM temperature
become larger than their virial temperature. Hence they cannot retain
efficiently their gas content and they lose their identity. We note
that neglecting the clumping of the gas would lead to a higher
reionization redshift $z_{ri}=7.3$ since it would underestimate the
efficiency of recombination. The clumping factor $C_n$ displays a behaviour similar to $C_b$ but it is usually smaller since it does not include the deep halos which can cool. We display in the lower panel of
Fig.\ref{figclumpO03} the overdensity $(1+\Delta)_{IGM}(z)$
characteristic of most of the volume of the universe at redshift $z$
when seen on scale $R_d$. While structures form and the matter gets
embedded within overdensities which occupy a decreasing fraction of
the volume (when seen at this scale $R_d$) the ``overdensity''
$(1+\Delta)_{IGM}$ which characterises the medium in-between these
objects (halos or filaments) declines. The density contrast $(1+\Delta)_n$ which corresponds to the IGM and shallow potential wells which do not form stars (but constitute Lyman-$\alpha$ clouds) decreases more slowly since it only excludes the high virial temperature halos with $T>T_{cool}$.

\begin{figure}[htb]

\centerline {\epsfxsize=8 cm \epsfysize=5.5 cm \epsfbox{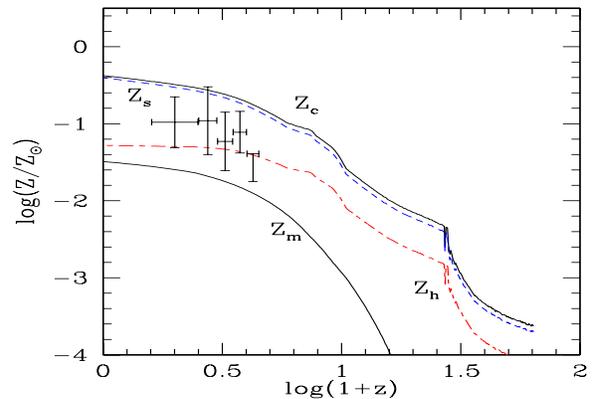}}

\caption{The redshift evolution of the metallicities $Z_c$ 
(star-forming gas), $Z_s$ (stars), $Z_h$ (galactic halos) and $Z_m$
(matter average). The data points are from Pettini et al.(1997) for
the zinc metallicity of damped Lyman-$\alpha$ systems.}
\label{figmetO03}

\end{figure}

We present in Fig.\ref{figmetO03} the redshift evolution of the mean
metallicities (in units of solar metallicity) $Z_c$ (within the
star-forming gas located in the inner parts of galaxies, upper solid
line), $Z_s$ (within stars, upper dashed line) and $Z_h$ (within
galactic halos, lower dashed line). We use the mass average over the
various galactic halos:
\beq
Z = \frac{ \int Z(M) \; \mu_g(M) \; \frac{dM}{M} } { \int \mu_g(M) \;
\frac{dM}{M} }
\eeq
where $\mu_g(M) dM/M$ is the galaxy mass function defined as in
(\ref{muh}). The reader is referred to VS II for a detailed
description of these metallicities (note that we only consider here
the abundance of Oxygen or any other element that is mainly produced
by SN II since we did not include SN I in our model). The lower solid
line corresponds to a ``matter averaged'' metallicity $Z_m$ defined by
$Z_m = F_{cool} Z_h$. Thus, although we do not include explicitly in
our model any contamination of the IGM by heavy elements produced
within galaxies, $Z_m$ defined in this way provides an upper bound for
the mean IGM metallicity (corresponding to very efficient mixing). If
galaxies do not eject metals very deeply within the IGM its
metallicity could be much smaller. The mean metallicity of
Lyman-$\alpha$ clouds associated to galactic halos (limit or damped
systems) is $Z_h$. Our results agree well with observations by Pettini
et al.(1997) for damped Lyman-$\alpha$ systems. Note that there is in
fact a non-negligible spread in metallicity over the various
halos.

\subsection{Consequence for the CMB radiation}

\begin{figure}[htb]

\centerline {\epsfxsize=8 cm \epsfysize=11.5 cm \epsfbox{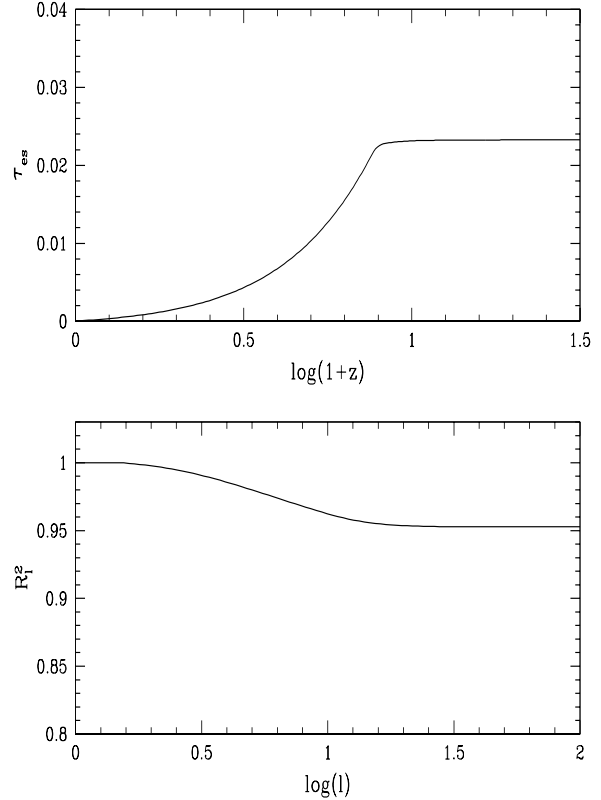}}

\caption{Upper panel: the optical depth $\tau_{es}$ for electron scattering. 
Lower panel: damping factor $R_l^2$ for the CMB power-spectrum.}
\label{figCMBO03}

\end{figure}

After reionization, CMB photons may be scattered by electrons present
in the gas. We write the corresponding Thomson opacity up to a
redshift $z$ as:
\beq
\tau_{es}(z) = \int_0^z c \frac{dt}{dz} \; \sigma_T \; \overline{n}_e(z)
\eeq
where $\overline{n}_e(z)$ is the mean electron number density at
redshift $z$. We take:
\beq
\overline{n}_e(z) = (1-Y) \; \frac{\Omega_b}{\Omega_0} \; \frac{\rhoa(z)}{m_p}
 \; \left( \frac{n_e}{n_H} \right)_{IGM}
\eeq
which means that we use the same electron fraction in clouds as for
the IGM (note that we calculate the IGM electron number density
together with the ionization state of hydrogen and helium). Then, CMB
anisotropies are damped on angular scales smaller than the angle
subtended by the horizon at reionization. We use the analytic fit
given by Hu \& White (1997) to obtain the damping factor $R_l^2$ of
the CMB power-spectrum $C_l$ from the optical depth $\tau_{es}$. The
results are shown in Fig.\ref{figCMBO03}.

\begin{figure}[htb]

\centerline {\epsfxsize=8 cm \epsfysize=5.5 cm \epsfbox{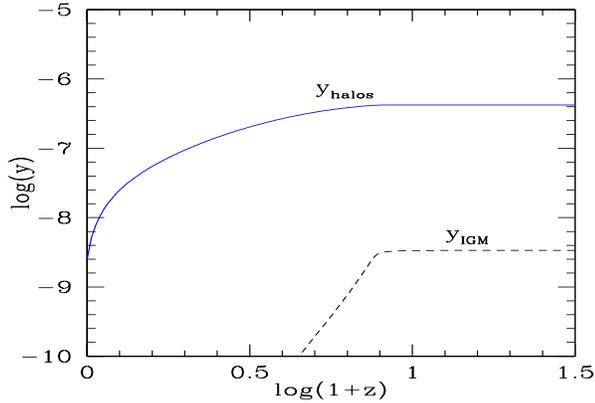}}

\caption{The Compton parameter $y$ up to redshift $z$ describing the 
Sunyaev-Zeldovich effect from the IGM (dashed line) and virialized
halos (solid line).}
\label{figycompO03}

\end{figure}

We can check that the total opacity $\tau_{es} \simeq 0.023$ is quite
small because reionization occurs rather late at $z_{ri} \simeq
6.8$. This also implies that the damping factor $R_l^2$ remains close
to unity: $R_l^2 \simeq 0.95$ for large $l$. Another distortion of the
CMB radiation is the Sunyaev-Zeldovich effect which transfers photons
from the Rayleigh-Jeans part of the spectrum to the Wien tail when
they are scattered by hot electrons. The magnitude of this
perturbation is conveniently described by the Compton parameter $y$:
\beq
y(z) = \int_0^z c \frac{dt}{dz} dz \; \sigma_T \; n_e \; \frac{kT}{m_e c^2}
\label{yComp}
\eeq
We can first consider the contribution of the IGM gas, using in
(\ref{yComp}) the temperature and the electronic density of this
uniform component. Then, we estimate the distortion due to the hot gas
embedded within virialized objects. We can write this latter
contribution as:
\beq
y_{halos} = \int dy_{IGM} \; \frac{1}{(1+\Delta)_{IGM}} \; F_{vir} 
\; \frac{T_{mvir}}{T_{IGM}}
\eeq
where we used the same electronic fraction for halos and the IGM. The
temperature $T_{mvir}$ is the ``mass averaged'' temperature of
virialized objects while $F_{vir}$ is the mass fraction within
collapsed objects displayed in Fig.\ref{figFracO03}. The results are
presented in Fig.\ref{figycompO03}.

The Compton parameter $y_{IGM}$ due to the IGM first increases rather
fast with $z$ until reionization, together with $n_e$ and
$T_{IGM}$. After $z_{ri}$ it reaches a plateau and does not grow any
more since at these large redshifts the universe is almost exactly
neutral. The contribution $y_{halos}$ of virialized objects is much
larger at low $z$ than $y_{IGM}$ since the temperature of these
collapsed halos is much higher than $T_{IGM}$, as shown in
Fig.\ref{figTO03}. We can note however that $y_{halos}$ grows much
slower and becomes close to its asymptotic value earlier than
$y_{IGM}$. This is due to the fact that the characteristic temperature
of virialized halos declines at larger $z$, see Fig.\ref{figTO03}, and
the mass fraction they contain also decreases (while the IGM undergoes
the opposite trends). A more detailed description of the
Sunyaev-Zeldovich effect due to clusters, and its fluctuations (which
in fact have the same magnitude as the mean), will be presented in a
future article.

\subsection{Contributions of quasars and stars}
\label{Contributions of quasars and stars}

\begin{figure}[htb]

\centerline {\epsfxsize=8 cm \epsfysize=11.5 cm \epsfbox{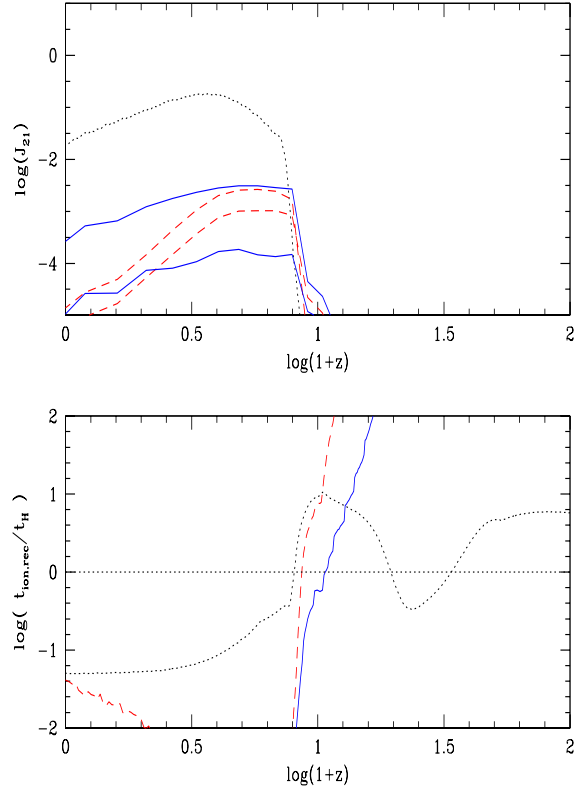}}

\caption{Upper panel: the redshift evolution of the ``instantaneous'' 
UV fluxes $J_{HI}^i$ and $J_{HeI}^i$ due to stars (solid lines)
and quasars (dashed lines). The dotted line is the UV flux $J_{21}$
shown in Fig.\ref{figJO03}. Lower panel: the ``instantaneous''
ionization times $t_{ion;s,Q}^i$ due to stars (solid line) and
quasars (dashed line). The dotted curve is the recombination time as
in Fig.\ref{figtionrecO03} while the horizontal solid line is the
Hubble time $t_H$.}
\label{figSourcesO03}

\end{figure}

In our model the radiation which reheats and reionizes the universe
comes from both quasars and stars. At large frequencies $\nu > 24.6$
eV most of the UV flux is emitted by quasars so that stars play no
role in the helium ionization. However, at lower frequencies both
contributions have roughly the same magnitude. We present in
Fig.\ref{figSourcesO03} the redshift evolution of the radiative output
due to stars and quasars.

Thus, we define the ``instantaneous'' radiation fields:
\beq
J_{\nu;s,Q}^i = \frac{1}{10} \; t_H \; S_{\nu;s,Q}
\eeq
where $t_H(z)$ is the Hubble time, see (\ref{Jnu}). From these
quantities we define the averages $J_{HI;s,Q}^i(z)$ and
$J_{HeI;s,Q}^i(z)$ as in (\ref{J21}). This provides a measure of the
radiative output above the ionization thresholds $\nu_{HI}$ and
$\nu_{HeI}$ due to stars and quasars. We can also derive the
ionization times $t_{ion;s,Q}^i$ as in (\ref{tion}). We can see in the
upper panel that at reionization $z \sim z_{ri}$ the contributions to
HI ionizing radiation from stars and quasars are of the same
order. However, since the quasar spectrum is much harder than stellar
radiation we have $J_{HeI;Q}^i \gg J_{HeI;s}^i$ so that quasars are
slightly more efficient at reheating and reionizing the universe (the
additional factor $(\nu-\nu_j)$ in (\ref{theat}) increases the weight
of high energy photons which also remain longer above the threshold
$\nu_{HI}$ while being redshifted). At low $z$ we can see that the
quasar radiative output declines much faster than the stellar source
term. Of course, this is due to the sharp drop at low redshifts of the
quasar luminosity function. This decrease of $S_{\nu Q}$ as compared
to $S_{\nu s}$ comes from two effects: i) as time increases the
``creation time-scale'' of halos of the relevant masses $M \sim
10^{12} M_{\odot}$ (through merging of smaller sub-units, measured by
$t_Q/t_M$) grows and ii) there is less gas available to fuel the
quasars (which even disappear) while old stars can still provide a
non-negligible luminosity source for galaxies. We can note in the
upper panel that the redshift evolution of the background radiation
field $J_{21}(z)$ actually produced by stars and quasars does not
exactly follow the ``instantaneous'' quantities $J_{HI;s,Q}^i(z)$
since one must take into account the expansion of the universe and
deviations from equilibrium. This explains the slower increase of
$J_{21}$ at $z \sim z_{ri}$ as well as the relatively low approximate
ionization times $t_{ion;s,Q}^i$ at this epoch.

\section{Critical universe}
\label{Critical universe}

We now consider the case of a critical universe $\Omega=1$ with a CDM
power-spectrum (Davis et al.1985) normalized to $\sigma_8=0.5$. We also choose
$\Omega_b=0.04$ and $H_0=60$ km/s/Mpc. Thus, as in the previous case
of an open universe our model is consistent with the studies
described in VS II and Valageas et al.(1999a).

\subsection{Quasar luminosity function}

\begin{figure}[htb]

\begin{picture}(230,430)

\epsfxsize=26 cm
\epsfysize=18 cm
\put(-28,-50){\epsfbox{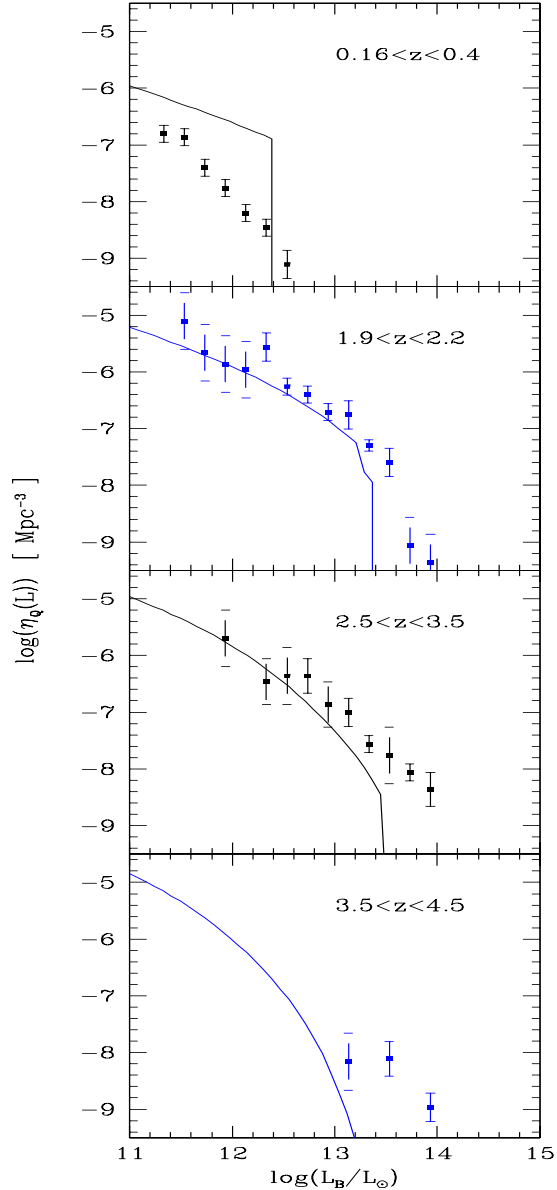}}

\end{picture}

\caption{The evolution with redshift of the B-band quasar luminosity 
function in comoving Mpc$^{-3}$, as in Fig.\ref{figquasO03}. The data
points are from Pei (1995).}
\label{figquasO1}

\end{figure}

We first present our results for the redshift evolution of the quasar
luminosity function in Fig.\ref{figquasO1}.

\begin{figure}[htb]

\centerline {\epsfxsize=8 cm \epsfysize=5.5 cm \epsfbox{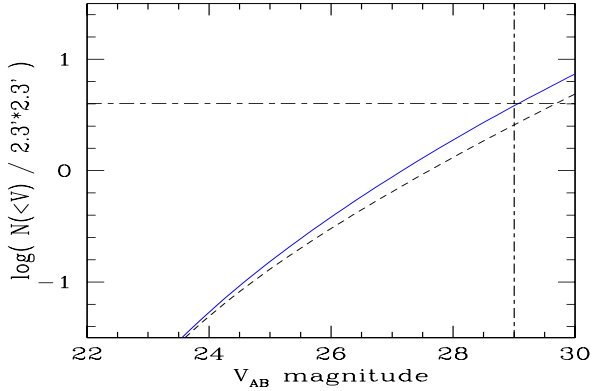}}

\caption{The quasar cumulative V-band number counts. The dashed line shows 
the counts of quasars with a magnitude brighter than $V$ located at
redshifts $3.5 < z < 4.5$ while the solid line corresponds to $3.5
< z < z_{ri}$.}
\label{figcountO1}

\end{figure}

We can check that our results are similar to those obtained
previously for an open universe and they again agree reasonably
with observations. This is not very surprising since we use the same
physical model so that we recover a similar behaviour. We present in
Fig.\ref{figcountO1} the quasar number counts we obtain from our
model.

We can see that our results are similar to those displayed previously
in Fig.\ref{figcountO03} and that our predictions are still marginally
consistent with the lack of observation in the HDF.

\subsection{Reheating}

\begin{figure}[htb]

\centerline {\epsfxsize=8 cm \epsfysize=5.5 cm \epsfbox{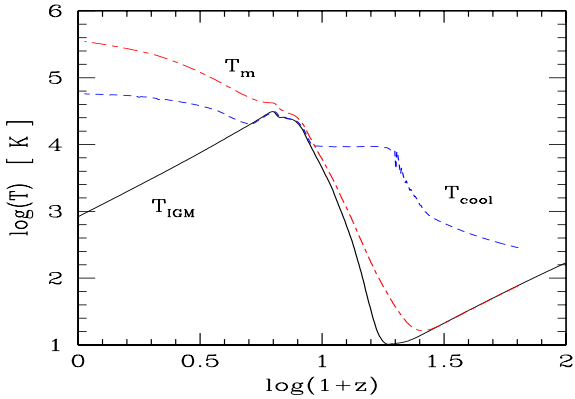}}

\caption{The redshift evolution of the IGM temperature $T_{IGM}$ 
(solid curve), the virial temperature $T_{cool}$ (upper dashed curve)
and the mass averaged temperature $T_m$ (lower dashed curve), as in
Fig.\ref{figTO03}.}
\label{figTO1}

\end{figure}

We present in Fig.\ref{figTO1} the redshift evolution of the IGM
temperature $T_{IGM}$, the virial temperature $T_{cool}$ of the
smallest objects which can cool at a given time, and the mass averaged
temperature $T_m$.

We can see that our results are again very close to those obtained for
an open universe. Indeed, the structure formation process is quite
similar and it must agree with the same observations (quasar and
galaxy luminosity functions, Lyman-$\alpha$ column density
distribution) at low $z$.

\subsection{Reionization}

\begin{figure}[htb]

\centerline {\epsfxsize=8 cm \epsfysize=11.5 cm \epsfbox{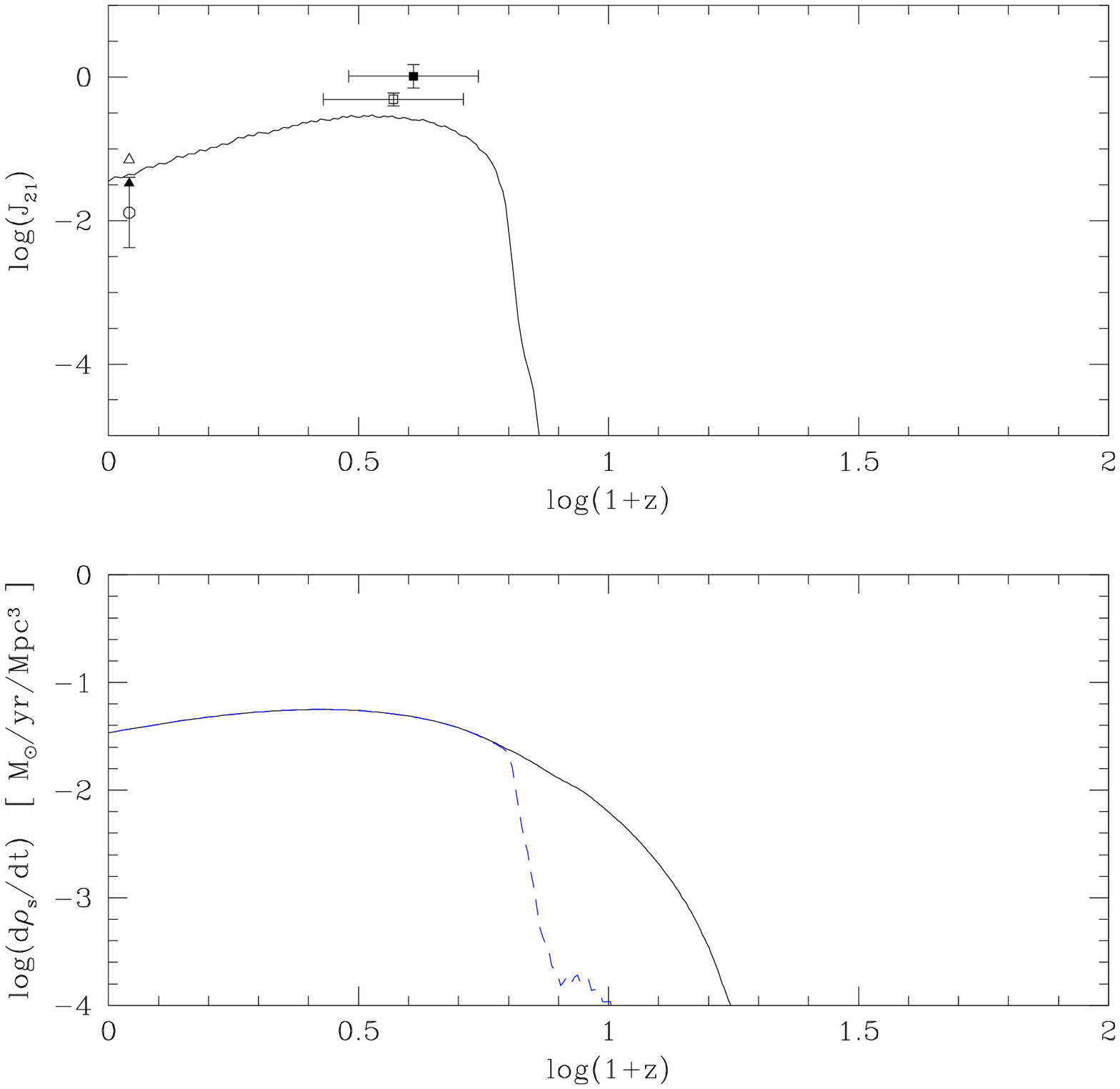}}

\caption{The redshift evolution of the UV flux $J_{21}$ (upper panel) and 
of the comoving star formation rate $d\rho_s/dt$ (lower panel) for the
case of an open universe. The dashed line in the lower panel shows the
effect of the absorption of high energy photons by the neutral
hydrogen present in the IGM and in Lyamn-$\alpha$ clouds.}
\label{figJO1}

\end{figure}

We display in Fig.\ref{figJO1} the redshift evolution of the
background radiation and the comoving star formation rate.

We can check that we recover the behaviour obtained previously for an open universe. However, the reionization redshift $z_{ri} = 5.6$
is smaller than previously. This is related to the lower normalization $\sigma_8$ of the power-spectrum as compared to the previous case. This leads to fewer bright quasars at high $z$ (compare Fig.\ref{figquasO1} and Fig.\ref{figquasO03}) and to a smaller radiative output.

We can also check that the hydrogen and helium reionization process is
close to our previous results (as for the reheating). Thus, for most
practical purposes both critical and open cosmologies allow reasonable
reheating and reionization histories which are very similar. In fact,
the uncertainties involved in the galaxy and quasar formation
processes are probably too large to favour significantly one of these
two possible scenarios (as compared to the other). However, both
models are consistent with present observations.

\section{Conclusion}

In this article we have described an analytic model for structure
formation in the universe which deals simultaneously with quasars,
galaxies and Lyman-$\alpha$ clouds, within the framework of a
hierarchical scenario. This allows us to study the reheating and
reionization history of the universe consistently with the properties
of these various classes of objects. We have shown that for both a
critical and an open universe our predictions agree reasonably well with
observations. However, as was noticed by Haiman et al.(1998) it
appears that the observational constraints on the quasar luminosity
function are already strong. Moreover, the Gunn-Peterson test for HeII
provides stringent additional constraints on the quasar contribution
to the UV radiation field and on the reionization redshift. Thus,
although our model in its simplest version (i.e. as described here,
with no additional cutoffs for the quasar multiplicity function) is
marginally consistent with the data, further observations of the helium
opacity and of the quasar number counts (e.g. with the NGST) could
provide tight constraints on such models where reionization is
produced by QSOs. On the other hand, since reionization occurs rather
late $z_{ri} \leq 7$ the damping of CMB anisotropies is quite small.

We can note that our predictions are similar to some results obtained by Gnedin
\& Ostriker (1997) with a numerical simulation (but for a different
cosmology). Moreover, the fact that our model agrees reasonably with
observations for Lyman-$\alpha$ clouds, galaxies, quasars and
constraints on the reionization process, strongly suggests that its
main characteristics are fairly realistic. Thus, it provides a simple
description of structure formation in the universe, from high
redshifts after recombination down to the present epoch.

\begin{acknowledgements}
This research has been supported in part by grants from NASA and NSF.
\end{acknowledgements}

\end{document}